\documentclass{aa}
\usepackage{graphicx}
\usepackage{txfonts}
\usepackage{physics}
\usepackage[utf8]{inputenc}
\usepackage[english]{babel}
\usepackage[T1]{fontenc}
\usepackage{lmodern}
\usepackage{microtype}
\usepackage{booktabs}
\usepackage{bm}
\usepackage{graphicx}
\usepackage{amsfonts}
\usepackage{amsmath}
\usepackage{amssymb}
\usepackage{listings}
\usepackage[hidelinks]{hyperref}
\usepackage{subfigure}
\usepackage{fancyhdr}
\usepackage{mathtools}
\usepackage{natbib}
\usepackage{comment}
\usepackage{color}
\usepackage{orcidlink}

\newcommand{\np}{\mathrm{NP}}
\newcommand{\msun}{\mathrm{M}_\odot}
\newcommand{\plpeak}{\textsc{PowerLaw+Peak}}

\bibpunct{(}{)}{;}{a}{}{,} 

\begin{document}

\titlerunning{Augmented mixture models for BH population studies}
\authorrunning{S.~Rinaldi}
\title{Expect the unexpected: Augmented mixture models for black-hole-population studies}
\author{Stefano~{Rinaldi}\,\orcidlink{0000-0001-5799-4155}\,\thanks{E-mail: stefano.rinaldi@uni-heidelberg.de}
                }
\institute{
    Institut für Theoretische Astrophysik, ZAH, Universität Heidelberg, Albert-Ueberle-Str.~2, 69120 Heidelberg, Germany
    }
\date{Received \today; accepted XXX}
\abstract
{Black-hole-population studies are currently performed either using astrophysically motivated models (informed but rigid in their functional forms) or via non-parametric methods (flexible, but not directly interpretable).}
{In this paper, we present a statistical framework to complement the predictive power of astrophysically motivated models with the flexibility of non-parametric methods.}
{Our method makes use of the Dirichlet distribution to robustly infer the relative weights of different models as well as of the Gibbs sampling approach to efficiently explore the parameter space.}
{After having validated our approach using simulated data, we applied this method to the binary black-hole mergers observed during the first three observing runs of the LIGO-Virgo-KAGRA collaboration using both phenomenological and astrophysical models as parametric models, finding results in agreement with the currently available literature.}
{}

\keywords{Methods: statistical -- gravitational waves -- stars: black holes}

\maketitle

\section{Introduction}
With 69 gravitational waves (GWs) from binary black-hole (BBH) mergers detected by the LIGO \citep{ligodetector:2015}, Virgo \citep{virgodetector:2015}, and KAGRA \citep{kagradetector:2021} collaboration (LVK) at the end of the third observing run (O3) \citep{gwtc3:2023} and 84 events newly added by GWTC-4.0 \citep{gwtc4:2025}, BBH population studies are now a prime tool for astrophysicists when it comes to investigating the physics of massive-binary evolution. The characterisation of the astrophysical BH distribution requires a profound understanding of the details of massive-star evolution such as the mass loss due to stellar winds \citep{kruckow:2024,romagnolo:2024,vink:2024,merritt:2025,vanson:2025}, its metallicity dependence \citep{belczynski:2016,hirschi:2025}, and the effect of the pair-instability supernova process \citep{mapelli:2020,vink:2021}. At the same time, it is also crucial to model the possible formation channels for compact binaries: the proposed models can be broadly divided into two classes: isolated evolution and dynamical formation. The isolated-evolution scenario considers the possibility of the progenitors of the two compact objects already being part of a binary system during the stellar stage: in this case, for example, the mass transfer (either stable or unstable) can tamper with the mass ratio of the binary \citep{roepke:2023} and thus leave an imprint on the resulting BH distribution \citep{marchant:2021,gallegos-garcia:2021,willcox:2023}. Conversely, dynamically assembled systems are binaries where the components are already brought together at the stage of compact objects as a result of dynamical interactions happening in dense environments, as is the case of three-body encounters, dynamical captures, and star clusters \citep{ziosi:2014,rodriguez:2016,kremer:2020,banerjee:2022}. For a comprehensive review of the available models, we refer the reader to \citet{mapelli:2020:review} and \citet{mandel:2022}. Due to the significant complexity of the astrophysical processes, the distribution induced by the aforementioned models cannot be expressed in terms of simple functions: the modelling community relies on numerical methods capable of producing synthetic catalogues of merging BBHs. For the same reason, developing an all-encompassing model accounting for all possible physical processes and formation channels is considered a titanic task.

Despite the effort put into accelerating and optimising population-synthesis codes, these algorithms are not yet fast enough to be embedded in the MCMC methods commonly used to analyse the GW data -- with the notable exception of machine-learning-enhanced approaches \citep[e.g.][]{colloms:2025}. The available literature often makes use of phenomenological parametrised models inspired by the expected features of the BH distribution: among others we mention the \textsc{PowerLaw+Peak} model \citep{astrodistGWTC3:2023}, describing the primary mass distribution as a weighted superposition of a tapered power-law and a Gaussian distribution, as well as its updated version \textsc{Broken Power Law + 2 Peaks} \citep{astrodistGWTC4:2025}, currently favoured by the data. These models, albeit simplified with respect to the fully fledged astrophysical models, are still capable of providing insights into the processes at play behind the observed BHs. This is the main approach employed by the LVK collaboration \citep{astrodistGWTC1:2019,astrodistGWTC2:2021,astrodistGWTC3:2023,astrodistGWTC4:2025} and many authors \citep[e.g.][]{fishbach:2017,talbot:2018,farah:2023,gennari:2025,rinaldi:2025:features}. For a comprehensive review, please see \citet{callister:2024:review}.
Due to the fact that the functional forms are merely inspired by the astrophysical models and do not have any direct connection with the physical processes at play, the possibility of assuming a functional form that does not encompass the true underlying distribution, and thus biasing the analysis, cannot be neglected entirely. Moreover, new, previously unforeseen features in the spectrum must be added by hand when needed.

To ensure the robustness of the population analysis in this respect, a complementary approach has been developed based on the concept of non-parametric methods. These, despite the potentially misleading name, are models that employ a functional form with a countably infinite number of parameters that can approximate arbitrary probability densities: in other words, a base for the space of normalised functions. These models have the advantage of being solely data-driven and not encoding any previous belief concerning the expected shape of the distribution, making them useful tools to discover new features in the BH spectrum. Among the ones developed within the GW community, we mention autoregressive processes \citep{callister:2024}, Dirichlet process Gaussian mixture models \citep{rinaldi:2022:hdpgmm}, Gaussian processes \citep{li:2021}, reversible-jump Markov chain Monte Carlo processes \citep{toubiana:2023}, and binned approaches \citep{ray:2023,heinzel:2024}. The flexibility of such models comes with the cost that the resulting reconstructed probability density is merely a description of the underlying distribution, not immediately interpretable in terms of astrophysical processes -- although in \citet{rinaldi:2025:np2p} we propose a way to circumvent this limitation.

In summary, we can broadly categorise the models employed as either `informed but rigid' (astrophysical and parametrised) or `flexible but not interpretable' (non-parametric), with somewhat complementary pros and cons. Several works employ a semi-parametric approach, in which the model is defined as the product of a parametric and a non-parametric distribution, each applied on a disjoint partition of the binary parameter space (e.g. a non-parametric model for the masses and a parametric model to describe the redshift and spin parameters). 
What is currently missing, to the best of our knowledge, is a way of bringing the two categories together to describe the same binary parameters in a `flexible and informed' way: in this work, we present the augmented mixture model (AMM), a weighted superposition of parametric (or astrophysical) and non-parametric models designed to infer the BH distribution accounting for the possible presence of unforeseen features in the spectrum while retaining the interpretability offered by the informed models.

The paper is organised as follows: in Section~\ref{sec:framework}, we summarise the statistical framework used to analyse GW data; Section~\ref{sec:amm} introduces the AMM, along with an outline of the algorithmic implementation; Section~\ref{sec:simulations} demonstrates the applicability of our method using synthetic data, whereas Section~\ref{sec:o3data} applies the AMM to the GW events detected during the third LVK observing run; lastly, Section~\ref{sec:conclusions} summarises our findings.

\section{Summary of statistical framework}\label{sec:framework}
In this work, we made use of the statistical framework presented in \citet{mandel:2018} and \citet{vitale:2022}, which is briefly recapped in this Section. Concerning the notation, we denote the data associated with the $N$ observed GW events included in the catalogue with $\mathbf{d} = \{d_1,\ldots d_N\}.$ Each GW signal is described by a set of parameters, $\theta$ (e.g. masses and spins of the binary components, distance, sky position, etc.). The fact that a specific GW event is detectable is denoted with $\mathbb{D}$. The astrophysical probability distribution is denoted with $p(\theta|\Lambda)$, where $\Lambda$ is the set of parameters used to describe the astrophysical distribution. Albeit not used in this section, in the specific case of non-parametric methods we denote the (infinitely many) parameters with $\Theta$.

Following \citet{mandel:2018}, the likelihood reads
\begin{equation}\label{eq:initial_likelihood}
p(\mathbf{d}|\Lambda,\mathbb{D}) = \prod_i^N p(d_i|\Lambda,\mathbb{D}) = \prod_i^N \int p(d_i|\theta_i,\mathbb{D})p(\theta_i|\Lambda,\mathbb{D})\dd \theta_i\,,
\end{equation}
having assumed that the events are independent and identically distributed. Making use of Bayes' theorem, the integrand can be refactored as 
\begin{equation}
\frac{p(\mathbb{D}|d_i,\theta_i)p(d_i|\theta_i)}{p(\mathbb{D}|\theta_i)}\frac{p(\mathbb{D}|\theta_i)p(\theta_i|\Lambda)}{p(\mathbb{D}|\Lambda)} = \frac{p(d_i|\theta_i)p(\theta_i|\Lambda)}{p(\mathbb{D}|\Lambda)}\,.
\end{equation}
The assumption here is that the detectability of an observed event is, by definition, equal to 1. The denominator is the detectability fraction,
\begin{equation}\label{eq:detectability_def}
p(\mathbb{D}|\Lambda) = \int p(\mathbb{D}|\theta)p(\theta|\Lambda)\dd \theta \equiv \xi(\Lambda)\,,
\end{equation}
and it is usually estimated via Monte Carlo integration using a set of simulated signals to marginalise over noise realisations.
Making use of Bayes' theorem on $p(d_i|\theta_i)$, Eq.~\eqref{eq:initial_likelihood} becomes
\begin{equation}\label{eq:std_likelihood}
p(\mathbf{d}|\Lambda, \mathbb{D}) = \prod p(d_i|\Lambda,\mathbb{D}) =  \prod_i^N \frac{p(d_i)}{\xi(\Lambda)}\int \frac{p(\theta_i|d_i)p(\theta_i|\Lambda)}{p(\theta_i)}\dd\theta_i \,.
\end{equation}
In the specific case in which the astrophysical model is a weighted superposition of $M$ models, whose parameters we denote with $\boldsymbol{\Lambda} = \{\Lambda_1, \ldots,\Lambda_M\}$,
\begin{equation}\label{eq:parametricmix}
p(\theta_i|\boldsymbol{\Lambda}) = \sum_j^M w_j p_j(\theta_i|\Lambda_j)\qq{with} \sum_j^Mw_j = 1\,,
\end{equation}
the likelihood takes the simple form of a superposition of likelihoods where the relative weights account for the different detectability fractions \citep{rinaldi:2025:features}:
\begin{multline}\label{eq:mixture_likelihood}
p(\mathbf{d}|\boldsymbol{\Lambda},\mathbb{D}) = \prod_i^N \sum_j^M \frac{w_j \xi_j(\Lambda_j)}{\sum_n w_n\xi_n(\Lambda_n)} p_j(d_i|\Lambda_j,\mathbb{D}), \\ \equiv \prod_i^N\sum_j^M \phi_j p_j(d_i|\Lambda_j,\mathbb{D})\,.
\end{multline}
Here, $p_j(d_i|\Lambda_j,\mathbb{D})$ refers to the likelihood defined in Eq.~\eqref{eq:std_likelihood} evaluated using the $j$-th model, $p_j(\theta|\Lambda_j)$; $\boldsymbol{\phi} \equiv \{\phi_1,\ldots\phi_M\} $ are the `observed' mixture fractions -- namely, the fraction of events that are generated from the corresponding mixture component after applying selection effects; and $\mathbf{w} \equiv \{w_1,\ldots,w_M\}$ denotes the `intrinsic' mixture fractions (as above, but before the application of selection effects).

\section{The augmented mixture model}\label{sec:amm}

In the previous section, we did not make any assumption regarding the specific models $p_j(\theta|\Lambda_j)$. In what follows, we consider a mixture of parametric models, plus one non-parametric model whose parameters are denoted by $\Theta$. In this case, differently from Eq.~\eqref{eq:parametricmix}, we require that $\Sigma_j w_j \leq 1$ to account for the presence of the additional non-parametric channel. Next,
\begin{equation}\label{eq:amm_definition}
p(\theta|\boldsymbol{\Lambda},\Theta) = \sum_j^M w_j p_j(\theta|\Lambda_j)\, +\, (1-\Sigma_j w_j) \np(\theta|\Theta)\,;
\end{equation}
here, we denote the non-parametric model with $\np(\theta|\Theta),$ as opposed to the parametric models, $p_j(\theta|\Lambda_j)$. We refer to this superposition as the `augmented mixture model' (AMM).

This specific choice, namely including a non-parametric component in the mixture, addresses the possibility that the parametric models might not capture all the features that are present in the underlying distribution encoded in the data; observations that are unlikely to be explained by the available parametric models -- either because they come from a region with little support or because there is an unforeseen overabundance of detections -- can be captured by the non-parametric channel, acting in this case as a sort of `additional storage', where the data that do not fit the analytical predictions can be collected. This ensures that only observations that are consistent with the functional form of the specific parametric model, $j,$ are considered while estimating its parameters, $\Lambda_j$, thus preventing mismodelling biases in the inferred posterior distribution, $p(\Lambda_j|\mathbf{d},\mathbb{D})$. In the same fashion, the non-parametric reconstruction was obtained making use of only the data that are unlikely to be explained by the available physically informed models, and thus describing only the features that are yet to be accounted for.
The property of the non-parametric model of being able to -- at least in principle -- approximate arbitrary probability densities translates to the AMM, making this model potentially overcomplete: this means that there might be more than one arbitrarily precise representation of the underlying data. This might be the case, for example, when a parametric model already including the features required to describe the data is augmented with a non-parametric model: both the case in which all the data are explained by the parametric model and the one where all the observations are captured by the non-parametric channel are precise descriptions of the underlying distribution. 
For the same reason, it is in principle possible for the non-parametric channel to `take over' the reconstruction and explain the entirety of the data alone, even if the parametric model could, in principle, account for a subset of the observations. This, however, is not expected to happen, due to the parametric model carrying more information about the expected shape of the distribution and thus being a priori favoured in certain areas of the parameter space with respect to the completely agnostic non-parametric method.

In the remainder of this section, we illustrate our algorithmic implementation of the AMM. In general, the joint parameter space $(\boldsymbol{\Lambda},\Theta)$ can be explored using a variety of techniques, mainly depending on the specific non-parametric model used; here, we make use of the collapsed Gibbs sampling approach and (H)DPGMM \citep{rinaldi:2022:hdpgmm} as the non-parametric method.

\subsection{Summary of (H)DPGMM and FIGARO}\label{sec:nonpar_summary}
We now recap the key aspects of both (H)DPGMM and its associated sampling scheme. For the full derivation, we refer the reader to \citet{rinaldi:2022:hdpgmm}.

The model we use, (H)DPGMM, is a non-parametric model based on the Gaussian mixture model (GMM), a potentially infinite weighted sum of multivariate Gaussian distributions able to approximate arbitrary probability densities \citep{nguyen:2020}:
\begin{equation}
        p(\theta) \simeq \np(\theta|\Theta) = \sum_{k = 1}^\infty \lambda_k \mathcal{N}(\theta|\mu_k,\sigma_k)\,.
\end{equation}
Here, the parameters are the weights $\boldsymbol{\lambda} \equiv \{\lambda_1,\lambda_2,\ldots\}$ with $\Sigma_k \lambda_k = 1$, the mean vectors $\boldsymbol{\mu} \equiv \{ \mu_1, \mu_2, \ldots\},$ and the covariance matrices $\boldsymbol{\sigma} \equiv \{\sigma_1, \sigma_2, \ldots\}$; they are collectively denoted with $\Theta = \{\boldsymbol{\lambda},\boldsymbol{\mu},\boldsymbol{\sigma}\}$. 
In \citet{rinaldi:2022:hdpgmm}, we introduced this non-parametric model as well as a scheme based on the collapsed Gibbs sampling approach to draw samples from the posterior distribution $p(\Theta|\mathbf{d},\mathbb{D})$. This scheme is implemented in the \textsc{figaro}\footnote{Publicly available at \url{https://github.com/sterinaldi/FIGARO} and via \href{https://pypi.org/project/figaro/}{\texttt{pip}}.} code \citep{rinaldi:2024:figaro}.
The potentially infinite number of Gaussian components in the mixture is accounted for by making use of a Dirichlet process \citep{ferguson:1973} prior on the weights, $\boldsymbol{\lambda}$, controlled by its concentration parameter, $\alpha_\textsc{dp}$; despite the number of components in a specific realisation of the GMM always being finite given a finite number of observations, $\mathbf{d}$, this choice makes it possible to --for every new data point added to the pool-- compute both the probability of the new $d_{N+1}$ having been drawn from one of the already observed $K$  Gaussian components (meaning that at least one of the other data points has been drawn from each of these $K$ components), as well as the probability of the new $d_{N+1}$ having been drawn from one of the infinitely many (and equally probable) unobserved components. This effectively a new Gaussian component to the mixture when required by the available data. 

In particular, if we introduce a set of indicator variables, $\mathbf{z} = \{z_1, \ldots z_N\}$, where each indicator variable, $z_i = k,$ reads ``the data $d_i$ has been drawn from the $k$-th Gaussian component'', it is possible to compute the probability of $d_{N+1}$ being drawn from the component, $k$ \citep[][Eq.~28]{rinaldi:2022:hdpgmm}:
\begin{multline}\label{eq:marginal_z}
p(z_{N+1} = k| \mathbf{z},\mathbf{d},\mathbb{D},\alpha_\textsc{dp})\\ = \frac{1}{\mathcal{K}} \int p(\mathbf{d}_{z_i = k}|\mu_k,\sigma_k,\mathbb{D})p(\mu_k,\sigma_k) \dd \mu_k\dd \sigma_k\\ \times p(z_{N+1} = k|\mathbf{z},\alpha_\textsc{dp})\,,
\end{multline}
where $p(\mathbf{d}_{z_i = k}|\boldsymbol{\mu},\boldsymbol{\sigma},\mathbb{D})$ is the likelihood in Eq.~\eqref{eq:std_likelihood} evaluated considering only the events assigned to the $k$-th component and the new $d_{N+1}$ if $k$ is one of the previously observed $K$ Gaussian components, or only $d_{N+1}$ if $ k = K+1$; i.e. a new, previously unobserved component. The categorical probability distribution expressed by the second term on the right hand side reads \citep[][Eq.~25 and 26]{rinaldi:2022:hdpgmm}
\begin{equation}\label{eq:categorical}
p(z_{N+1} = k|\mathbf{z},\alpha_\textsc{dp}) = \begin{cases} \frac{\alpha_\textsc{dp}}{N+\alpha_\textsc{dp}} \qif{k = K+1} \\ \frac{n_k}{N+\alpha_\textsc{dp}} \qq{otherwise}\end{cases}\,.
\end{equation}
Here, $n_k$ denotes the number of events already associated with the component, $k$. Summing over all the possible components plus the new, unobserved one gives the normalisation constant, $\mathcal{K}$:
\begin{equation}\label{eq:normconst}
\mathcal{K} = \sum_{k=1}^{K+1} p(z_{N+1} = k | \mathbf{z},\mathbf{d},\mathbb{D}, \alpha_\textsc{dp})
.\end{equation}

The introduction of indicatory variables is particularly useful when it comes to sampling the posterior distribution $p(\Theta|\mathbf{d},\mathbb{D})$. Since all Gaussian components of the mixture are independent of each other, conditioning on $\mathbf{z}$ makes the parameter space partially separable:
\begin{multline}\label{eq:factorised_posterior}
p(\boldsymbol{\lambda},\boldsymbol{\mu},\boldsymbol{\sigma}|\mathbf{d},\mathbb{D},\mathbf{z},\alpha_\textsc{dp}) = p(\boldsymbol{\lambda}|\mathbb{D},\mathbf{z},\boldsymbol{\mu},\boldsymbol{\sigma},\alpha_\textsc{dp})\\ \times \prod_{k=1}^K p(\mu_k,\sigma_k|\mathbf{d}_{z_i = k},\mathbb{D})\,.
\end{multline}
The posterior distribution on $\boldsymbol{\lambda}$, under the assumption of a Dirichlet prior, reads
\begin{multline}\label{eq:lambda_posterior}
p(\boldsymbol{\lambda}|\mathbb{D},\mathbf{z},\boldsymbol{\mu},\boldsymbol{\sigma},\alpha_\textsc{dp}) = \int p(\boldsymbol{\lambda}|\boldsymbol{\phi},\mathbb{D},\boldsymbol{\mu},\boldsymbol{\sigma})p(\boldsymbol{\phi}|\mathbf{z},\alpha_\textsc{dp})\dd \boldsymbol{\phi} \\ = \int \prod_k \delta\qty(\lambda_k - \phi_k/\xi(\mu_k,\sigma_k))p(\boldsymbol{\phi}|\mathbf{z},\alpha_\textsc{dp}) \,,
\end{multline}
where we made use of the definitions given in Eq.~\eqref{eq:detectability_def} for $\xi(\mu,\sigma)$ and Eq.~\eqref{eq:mixture_likelihood} for $\boldsymbol{\phi}$, and
\begin{equation}\label{eq:phi_posterior}
p(\boldsymbol{\phi}|\mathbf{z},\alpha_\textsc{dp}) = \Gamma(N+\alpha_\textsc{dp}) \prod_{k=1}^K \frac{\phi_k^{(n_k+\alpha_\textsc{dp}/K) -1}}{\Gamma(n_k + \alpha_\textsc{dp}/K)}\,.
\end{equation}

This parameter space is explored by \textsc{figaro} using the collapsed Gibbs sampling approach. The Gibbs sampling scheme \citep{geman:1984,gelfand:1990,smith:1993} is useful in all these situations where directly sampling from the joint parameter space --in our case, $p(\Theta, \mathbf{z}|\mathbf{d}, \mathbb{D})$ -- is expensive or impossible, but sampling from the conditioned distributions $p(\mathbf{z}|\Theta, \mathbf{d}, \mathbb{D})$ and $p(\Theta|\mathbf{z}, \mathbf{d}, \mathbb{D})$ is simple. If one of the conditional probabilities can be efficiently marginalised over the conditioned variable, thus making the exploration of the parameter space even simpler, the scheme is referred to as collapsed Gibbs sampling \citep{liu:1994}; this is the case, for example, of $p(\mathbf{z}|\mathbf{d},\mathbb{D})$ in Eq.~\eqref{eq:marginal_z}. 

Operatively, \textsc{figaro} draws a single sample from the posterior distribution $p(\Theta|\mathbf{d},\mathbb{D})$ via the following steps.
\begin{enumerate}
        \item Draw a sample for $\mathbf{z}$:
        \begin{enumerate}
                \item Consider an empty mixture model, with no observations. Out of the $N$ available events, pick one at random and assign it to the first component. At this stage, $\mathbf{z} = \{z_1 = 1\}$.
                \item Randomly pick a new event from the remaining $N-1.$  Using Eq.~\eqref{eq:marginal_z} conditioned on the current value of $\mathbf{z}$, compute the probability of assigning this second event to the same component as the first one ($z_2 = 1$) or to a new one ($z_2 = 2$) and draw $z_2$ accordingly.
                \item Repeat this procedure until all the observations have been added to the mixture, which will eventually have $K$ active components. This will be the $\mathbf{z}$ sample.
        \end{enumerate}
        \item Draw a sample for $\Theta$ conditioned on the $\mathbf{z}$ sample:
        \begin{enumerate}
                \item For each of the $K$ active components of the mixture, draw a sample for $(\mu_k,\sigma_k)$ using each of the terms in the product of Eq.~\eqref{eq:factorised_posterior}. These will be the $\boldsymbol{\mu}$ and $\boldsymbol{\sigma}$ samples.
                \item From Eq.~\eqref{eq:phi_posterior}, draw a sample for $\boldsymbol{\phi}$.
                \item Using the $\boldsymbol{\mu}$ and $\boldsymbol{\sigma}$ samples to compute $\xi(\mu_k,\sigma_k)$, convert the $\boldsymbol{\phi}$ sample into a $\boldsymbol{\lambda}$ sample to obtain $\Theta = \{\boldsymbol{\lambda},\boldsymbol{\mu},\boldsymbol{\sigma}\}$.
        \end{enumerate}
\end{enumerate}
These steps can be repeated to produce as many samples for $\Theta$ as needed. The $\mathbf{z}$ samples, at this stage, are just a byproduct and can be discarded.

\subsection{Inferring the parameters of an AMM}
We now turn our attention to the problem of exploring the parameter space of the AMM, leveraging on the same sampling scheme used for the non-parametric method.
From a mathematical point of view, most of the derivation is identical to the one we just summarised for the non-parametric methods. The main difference is that now the number of components -- parametric models plus a non-parametric one -- is finite and equal to $M+1$. Therefore, the prior on the weights, $\mathbf{w,}$ is not a Dirichlet process, but its finite equivalent, the Dirichlet distribution. This distribution takes, as input, a vector of $M+1$ positive numbers $\boldsymbol{\gamma} = \{\gamma_1,\ldots,\gamma_{M+1}\}$ acting as a priori pseudo-counts for each channel. The choice in which $\gamma_i = 1\ \forall i$ corresponds to the uniform distribution on the $M$-dimensional simplex.

In particular, denoting the indicator variables for the augmented mixture model with $\boldsymbol{\zeta} = \{\zeta_1,\ldots,\zeta_N\},$
 Eq.~\eqref{eq:marginal_z} becomes
\begin{multline}\label{eq:parametric_predictive}
        p(\zeta_{N+1} = j | \boldsymbol{\zeta}, \mathbf{z}, \mathbf{d},\mathbb{D},\boldsymbol{\gamma}, \alpha_\textsc{dp})\\ = \frac{1}{\mathcal{C}}\int p_j(\mathbf{d}_{\zeta_i = j}|\Lambda_j,\mathbb{D})p(\Lambda_j) \dd \Lambda_j \\ \times p(\zeta_{N+1} = j|\boldsymbol{\zeta},\boldsymbol{\gamma}) \,,
\end{multline}
if $j < M+1$, thus corresponding to a parametric model, and
\begin{multline}\label{eq:nonpar_predictive}
        p(\zeta_{N+1} = M+1 | \boldsymbol{\zeta}, \mathbf{z}, \mathbf{d},\mathbb{D},\boldsymbol{\gamma},\alpha_\textsc{dp})\\ = \frac{1}{\mathcal{C}} \int p(\mathbf{d}_{\zeta_i = M+1}|\Theta,\mathbb{D}, \mathbf{z},\alpha_\textsc{dp})p(\Theta) \dd \Theta\\ \times  p(\zeta_{N+1} = M+1|\boldsymbol{\zeta},\boldsymbol{\gamma}) \,,
\end{multline}
when considering the non-parametric model. The integrand can be further broken down using $\mathbf{z}$:
\begin{multline}
        p(\mathbf{d}_{\zeta_i = M+1}|\Theta,\mathbb{D}, \mathbf{z},\alpha_\textsc{dp})\\ = \sum_{k=1}^{K+1}\int p(\mathbf{d}_{\zeta_i = M+1, z_i = k}|\mu_k,\sigma_k, \mathbb{D})p(\mu_k,\sigma_k)\dd \mu_k\dd\sigma_k\\ \times p(z_{N+1} = k|\mathbf{z},\alpha_\textsc{dp})\,.
\end{multline}
These are the same terms used in the non-parametric inference presented in the previous Subsection just with the additional condition of considering only the events actually associated with the non-parametric channel. The categorical probability distribution corresponding to the one in Eq.~\eqref{eq:categorical} reads, due to the finite number of mixture components,
\begin{equation}
        p(\zeta_{N+1} = j|\boldsymbol{\zeta},\boldsymbol{\gamma}) = \frac{n_j+\gamma_j}{N+\Sigma_n \gamma_n}\,.
\end{equation}
As in Eq.~\eqref{eq:normconst}, the normalisation constant $\mathcal{C}$ in Eqs.~\eqref{eq:parametric_predictive} and~\eqref{eq:nonpar_predictive} is simply the sum over all the possible models:
\begin{equation}
        \mathcal{C} = \sum_{j=1}^{M+1} p(\zeta_{N+1} = j | \boldsymbol{\zeta}, \mathbf{z}, \mathbf{d},\mathbb{D},\boldsymbol{\gamma}, \alpha_\textsc{dp})\,.
\end{equation}
Finally, the posterior distribution on $(\mathbf{w},\boldsymbol{\Lambda},\Theta)$ can be expressed as
\begin{multline}\label{eq:parametric_posterior}
p(\mathbf{w},\boldsymbol{\Lambda},\Theta|\mathbf{d},\mathbb{D},\boldsymbol{\zeta},\mathbf{z},\boldsymbol{\gamma},\alpha_\textsc{dp}) = p(\mathbf{w}|\boldsymbol{\Lambda},\Theta,\mathbb{D},\boldsymbol{\zeta},\boldsymbol{\gamma})\\ \times p(\Theta|\mathbf{d}_{\zeta_i = M+1},\mathbb{D},\mathbf{z}, \alpha_\textsc{dp})\prod_{j=1}^Mp(\Lambda_i|\mathbf{d}_{\zeta_i = j},\mathbb{D})\,,
\end{multline}
where $p(\Theta|\mathbf{d}_{\zeta_i = M+1},\mathbb{D},\mathbf{z}, \alpha_\textsc{dp})$ is given by Eq.~\eqref{eq:factorised_posterior} and
\begin{multline}
p(\mathbf{w}|\boldsymbol{\Lambda},\Theta,\mathbb{D},\boldsymbol{\zeta},\boldsymbol{\gamma}) = \int p(\mathbf{w}|\boldsymbol{\phi},\mathbb{D},\boldsymbol{\Lambda},\Theta) p(\boldsymbol{\phi}|\boldsymbol{\zeta},\boldsymbol{\gamma})\dd \boldsymbol{\phi}\\ = \int \delta(w_{M+1}-\phi_{M+1}/\xi(\Theta))\prod_{j=1}^M \delta(w_j-\phi_j/\xi_j(\Lambda_j))\\ \times p(\boldsymbol{\phi}|\boldsymbol{\zeta},\boldsymbol{\gamma})\dd \boldsymbol{\phi}\,.
\end{multline}
The posterior distribution for $\boldsymbol{\phi}$ is the same as Eq.~\eqref{eq:phi_posterior}:
\begin{equation}\label{eq:phi_dirichletdist}
p(\boldsymbol{\phi}|\boldsymbol{\zeta},\boldsymbol{\gamma}) = \Gamma(N + \Sigma_n\gamma_n)\prod_{j=1}^{M+1}\frac{w_j^{n_j+\gamma_j-1}}{\Gamma(n_j+\gamma_j)}\,.
\end{equation}
In this derivation, we tacitly assumed that the parametric models have completely disjoint sets of parameters. If this is not the case, the models sharing at least one parameter have to be inferred jointly.

With all these ingredients, we are able to draw samples from the posterior distribution $p(\boldsymbol{\Lambda},\Theta|\mathbf{d},\mathbb{D})$. The steps are listed below.
\begin{enumerate}
        \item Draw a sample for $\boldsymbol{\zeta}$ and $\mathbf{z}$.
        \begin{enumerate}
                \item Starting from an empty mixture, randomly select one of the available events and compute the probability of assigning it to one of the parametric models (Eq.~\eqref{eq:parametric_predictive}) or to the non-parametric component (Eq.~\eqref{eq:nonpar_predictive}); then, draw $\zeta_1$ accordingly. If $\zeta_1 = M+1$, also assign the event to one of the Gaussian components of the non-parametric model.
                \item Repeat the previous step adding all the $N-1$ remaining events in a random order, each time conditioning on the current values of $\boldsymbol{\zeta}$ and $\mathbf{z}$. Events for which $\zeta_i = M+1$ should also be used to update $\mathbf{z}$ following the procedure described in Section~\ref{sec:nonpar_summary}. This will produce a sample for $\boldsymbol{\zeta}$ and one for $\mathbf{z}$.
        \end{enumerate}
        \item Draw a sample for $\boldsymbol{\Lambda}$, $\Theta,$ and $\mathbf{w}$ conditioned on $\boldsymbol{\zeta}$ and $\mathbf{z}$.
        \begin{enumerate}
                \item For each parametric model, draw a sample for $\Lambda_j$ using a stochastic sampling scheme (e.g. Markov Chain Monte Carlo, MCMC) for every term of the product in Eq.~\eqref{eq:parametric_posterior}, producing a sample for $\boldsymbol{\Lambda}$.
                \item Draw a sample for $\Theta$ as in Section~\ref{sec:nonpar_summary}.
                \item From Eq.~\eqref{eq:phi_dirichletdist}, draw a sample for $\boldsymbol{\phi}$.
                \item Using the $\boldsymbol{\Lambda}$ and $\Theta$ sample to compute $\xi_j(\Lambda_j)$ and $\xi(\Theta)$, convert the $\boldsymbol{\phi}$ sample into a $\mathbf{w}$ sample.
        \end{enumerate}
\end{enumerate}
Again, this procedure can be iterated to produce multiple $(\mathbf{w},\boldsymbol{\Lambda},\Theta)$ samples. In this case, the auxiliary variable $\boldsymbol{\zeta}$ can be useful in determining which channel is more likely to have produced each of the available events. 

To accompany this derivation, we developed \textsc{anubis}, a Python code implementing the algorithm presented in this Section. \textsc{anubis} relies on \textsc{figaro} for the non-parametric inference and on \textsc{Eryn} \citep{foreman-mackey:2013,karnesis:2023,katz:2023} as an MCMC sampler. 
In our implementation, the integral in Eq.~\eqref{eq:parametric_predictive} is carried out using a Monte Carlo approximation with samples drawn from the prior $p(\Lambda_j)$:
\begin{multline}
    \int p_j(\mathbf{d}_{\zeta_i = j}|\Lambda_j,\mathbb{D})p(\Lambda_j) \dd \Lambda_j \\ \simeq \frac{1}{N_s} \sum_{\ell=1}^{N_s} p_j(\mathbf{d}_{\zeta_i = j}|\Lambda_{j,\ell},\mathbb{D})\bigg\rvert_{\Lambda_{j,\ell}\sim p(\Lambda_j).}
\end{multline}
A single sample out of the $N_s$ used in the Monte Carlo sum is also randomly selected at every iteration with probability proportional to $p_j(\mathbf{d}_{\zeta_i = j}|\Lambda_{j,\ell},\mathbb{D})$ to be used as starting point for the MCMC chain of step 2a. This ensures that the chain is already in a thermalised state -- shortening the overall algorithm runtime -- and that subsequent samples of $\Lambda_j$ are independent.
The code is publicly available\footnote{It can be found at \url{https://github.com/sterinaldi/ANUBIS} and also installed via \href{https://pypi.org/project/anubis/}{\texttt{pip}}} and is used in the analyses of the following sections.

\section{Simulated data}\label{sec:simulations}
To demonstrate the capability of AMMs to capture features in distributions, we applied the framework described in the previous section to two different simulated scenarios inspired by the BH distribution presented in \citet{astrodistGWTC3:2023}. The inferred values for the parameters are reported quoting the median and $68\%$ credible interval.

\subsection{One-dimensional distribution}\label{sec:1dsim}
\begin{figure*}
    \centering
    \subfigure[Full AMM reconstruction]{
    \includegraphics[width=0.93\columnwidth]{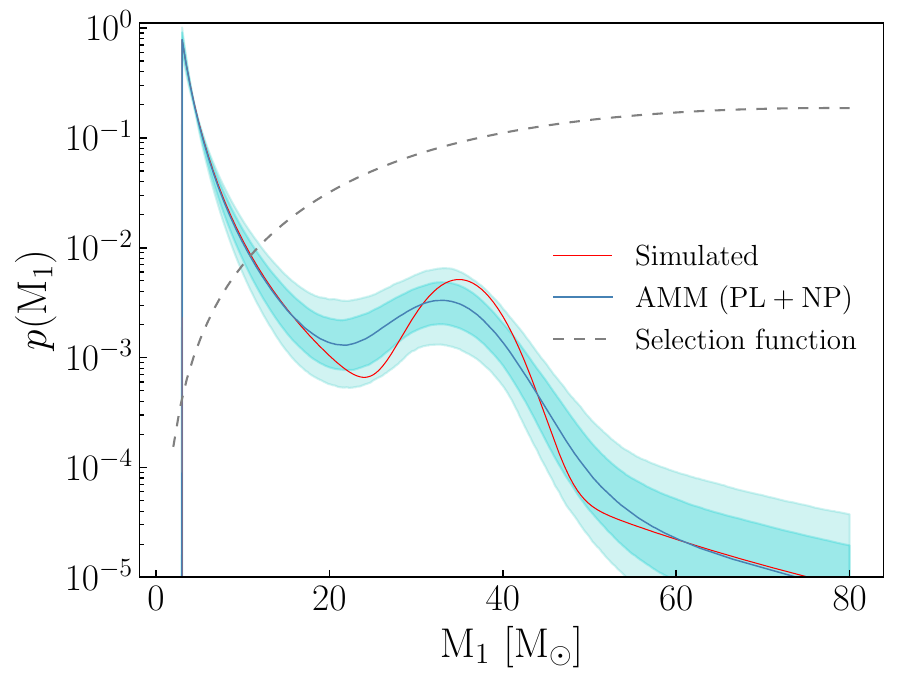}\label{fig:onedim_massdist}
    }
    \subfigure[Non-parametric reconstruction]{\includegraphics[width=0.93\columnwidth]{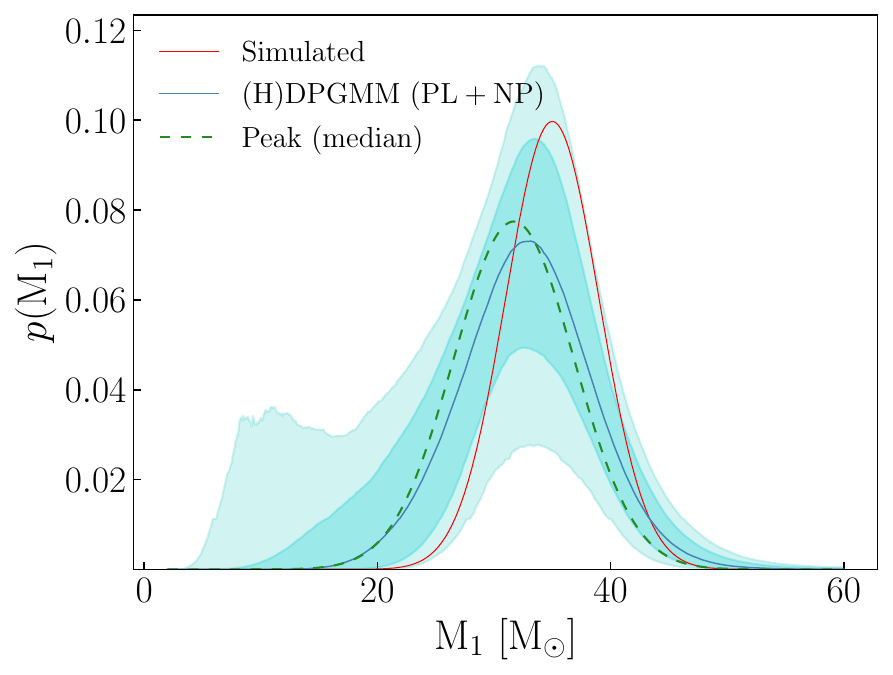}\label{fig:onedim_nonpar}
    }
    \caption{Inferred distribution for one-dimensional \plpeak~example presented in Section~\ref{sec:1dsim} in the PL+NP case. The solid blue line marks the median reconstruction, the shaded areas correspond to the $68\%$ and $90\%$ credible regions, and the solid red line shows the true underlying distribution. The grey dashed line in the left panel corresponds to the selection function used.}
\end{figure*}

The first, simplified example presented here is a one-dimensional distribution describing the BBH primary mass. The underlying distribution is assumed to be a truncated power-law (PL) distribution with the spectral index $\alpha$, bounded between between $M_\mathrm{min} = 3\ \msun$ and $M_\mathrm{max} = 80\ \msun$ (assumed known) and superimposed to a Gaussian distribution:
\begin{equation}
p(M|\alpha,\mu,\sigma,w) = w_\mathrm{PL}\mathrm{PL}(M|\alpha) + w_\mathrm{Peak}\mathcal{N}(M|\mu,\sigma)\,.
\end{equation}
The PL distribution is defined as
\begin{equation}
\mathrm{PL}(M|\alpha) = \begin{cases} \frac{(1-\alpha)M^{-\alpha}}{M_\mathrm{max}^{1-\alpha} - M_\mathrm{min}^{1-\alpha}} \qif M_\mathrm{min} < M < M_\mathrm{max} \\ \qq{}\ 0 \qq{}\ \,\qq{otherwise} \end{cases}\,.
\end{equation}
To generate the mock data presented in this section, we used $\alpha = 3.5$, $\mu = 35\ \msun$, $\sigma = 4\ \msun,$ and $w_\mathrm{Peak} = 0.05$. The selection function through which we filtered the data, reported in Figure~\ref{fig:onedim_massdist} as a dashed grey line, is modelled after the one presented in \citet{veske:2021}. From this distribution we draw 100 observed values $\boldsymbol{M} = \{M_1,\ldots, M_{100}\}$, and for each of these we simulated a posterior distribution $p(\theta_i|d_i)$ as a log-normal distribution as
\begin{equation}
p(\theta_i|d_i) = \mathcal{N}(\log M| \log m_i,\sigma_r)
,\end{equation}
where $\sigma_r = 0.15$ for all events and
\begin{equation}
m_i \sim \mathcal{N}(\log m| \log M_i,\sigma_r)\,.
\end{equation}
In this simplified example, we do not account for the correlation between detection efficiency and measurement uncertainty: we include this feature in the more realistic simulation presented in the next section. In what follows, we analyse these simulated data with different AMMs.

\begin{figure}
        \centering
        \includegraphics[width=\columnwidth]{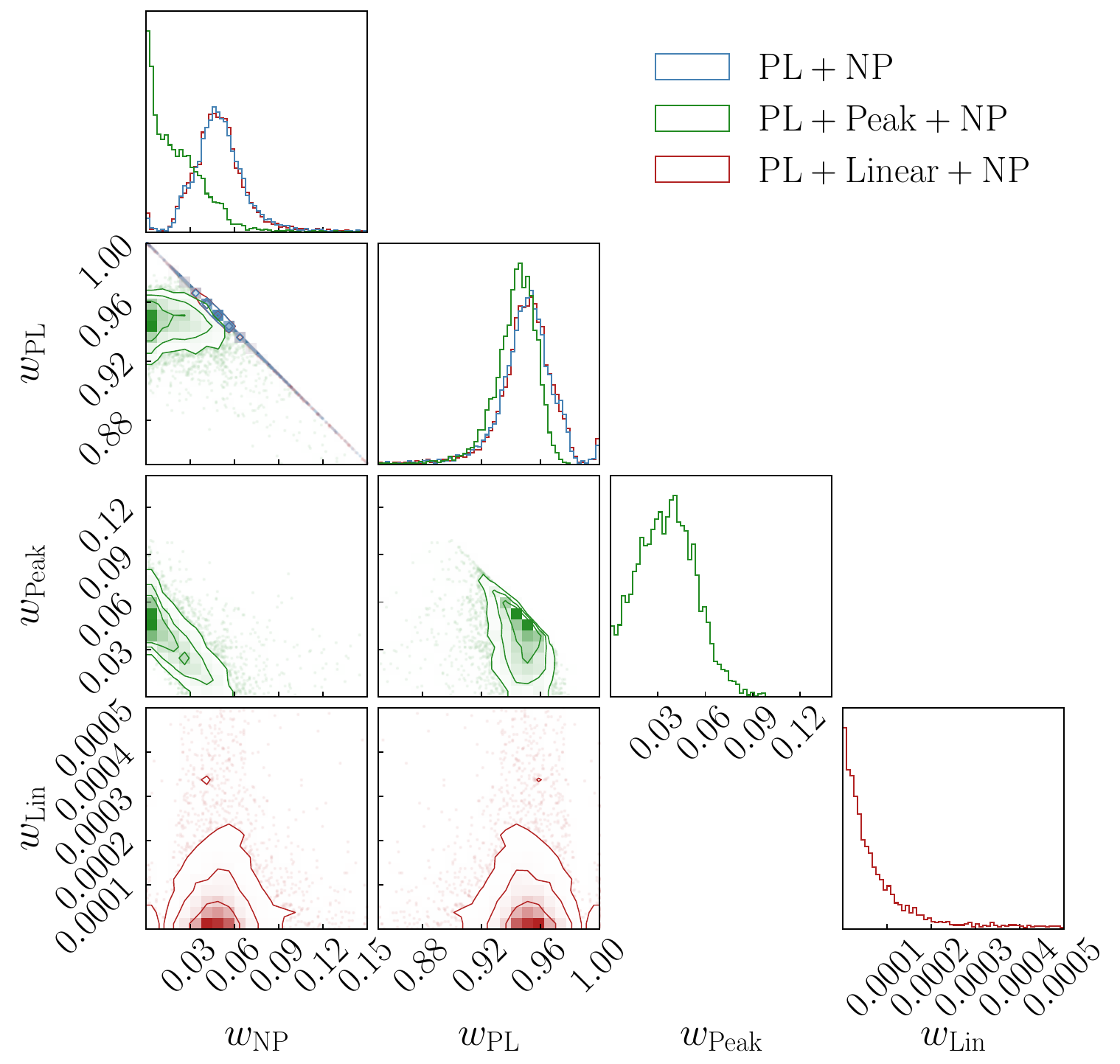}
        \caption{Inferred weights for different mixture models considered in Section~\ref{sec:1dsim}.}\label{fig:onedim_weights}
\end{figure}

\subsubsection{PL + NP}
In this first test, we assume a PL model augmented with the non-parametric channel (PL+NP), simulating the scenario where we have a good theoretical understanding of only one BBH formation channel. The free parameters are the PL index $\alpha$ and the relative weights $\mathbf{w} = \{w_\mathrm{NP}, w_\mathrm{PL}\}$. The inferred distribution is reported in Figure~\ref{fig:onedim_massdist}. The inferred value for the PL index is $\alpha = 3.5\pm0.4$, which is consistent with the simulated value. For comparison, if we do not include the non-parametric channel, we obtain $\alpha = 2.4\pm 0.1$ -- a value biased by the presence in the data of a feature not accounted for in the model. 

The non-parametric reconstruction, reported in Figure~\ref{fig:onedim_nonpar}, highlights the presence of a feature at around $30-35\ \msun$ consistent with the simulated Gaussian distribution. The weights $(w_\mathrm{NP},\ w_\mathrm{PL})$ are reported in Figure~\ref{fig:onedim_weights} (blue histograms). The inferred non-parametric weight is $w_\mathrm{NP} = 0.05^{+0.02}_{-0.01}$, in agreement with the simulated value $w_\mathrm{Peak} = 0.05$ that the non-parametric channel is expected to capture.

\subsubsection{PL + linear + NP}
Secondly, we show that the AMM is robust with respect to the inclusion of a `useless' component -- a channel not part of the underlying distribution -- in the mixture. In particular, we add a linear distribution bounded between the same $M_\mathrm{min}$ and $M_\mathrm{max}$ as the PL distribution:
\begin{equation}
\mathrm{Lin}(M) = \begin{cases} \frac{2M}{M_\mathrm{max}^2 - M_\mathrm{min}^2} \qif M_\mathrm{min} < M < M_\mathrm{max} \\ \qq{}\ 0 \qq{}\qq{otherwise} \end{cases}\,.
\end{equation}
The red histogram in Figure~\ref{fig:onedim_weights} shows how the inference is unaffected by the presence of an additional, unused channel: the weight associated with the linear distribution is found to be negligible and thus such channel does not contribute to the overall inference, leading to a reconstruction almost identical to the PL+NP case.

\subsubsection{PL + peak + NP}
The last case considered here is the one in which we have all the necessary components in the parametric mixture to describe the underlying distribution, thus making the non-parametric channel redundant: in particular, we include a PL distribution, a Gaussian peak, and a non-parametric model in the analysis. We find, for the parameters of this mixture, $\alpha = 3.7\pm 0.4$, $\mu = 32^{+3}_{-4}\ \msun$ and $\sigma = 5^{+2}_{-3}\ \msun$ -- all in agreement with the simulated values. Figure~\ref{fig:onedim_nonpar} reports, with a dashed green, the Gaussian distribution corresponding to the median inferred $\mu$ and $\sigma$: this is in good agreement with the non-parametric feature reconstructed in the PL+NP case. 

The green histograms in Figure~\ref{fig:onedim_weights} show the posterior distribution for the three weights $(w_\mathrm{NP},w_\mathrm{PL},w_\mathrm{Peak})$: in this case, the presence of additional, non-parametric features is disfavoured by the presence of the Gaussian channel ($w_\mathrm{NP} = 0.01^{+0.02}_{-0.01}$ and $w_\mathrm{Peak} = 0.04\pm0.02$). The suppression of the unused channel is, however, not as confident as it is in the previous PL+linear+NP case due to the flexibility of the non-parametric model. The mixture model being over-complete, it is always possible to associate some of the available events with the NP channel, thus resulting in a non-zero associated weight. Nonetheless, the shape of the posterior distribution for $w_\mathrm{NP}$ suggests that the parametric models alone are sufficient to explain the data.

\subsection{Three-dimensional distribution}\label{sec:3dsim}
We now consider a more realistic example modelled after the real LVK GW detections, including the primary mass $\mathrm{M}_1$, the mass ratio $q,$ and the redshift $z$ in the analysis. The primary mass follows a \plpeak~distribution (described in Appendix B1.b of \citealt{astrodistGWTC3:2023}) with parameters $\alpha = 3.5$, $\delta = 5\ \msun$, $\mathrm{M}_\mathrm{min} = 3\ \msun$, $\mu = 35\ \msun$, $\sigma = 4\ \msun,$ and $w_\mathrm{Peak} = 0.02$. The maximum mass, $\mathrm{M}_\mathrm{max} = 80\ \msun,$ is assumed known. The mass ratio is distributed according to a power law:
\begin{equation}
p(q|\beta) \propto q^\beta
,\end{equation}
with $\beta = 1.1,$ and the redshift is
\begin{equation}\label{eq:redshiftdist}
p(z|\kappa) \propto \frac{\mathrm{d}V}{\mathrm{d}z} (1+z)^{\kappa-1}
,\end{equation}
with $\kappa = 2$ between $z=0$ and $z = 2$. For all the other parameters (spins and extrinsic parameters), we used isotropic and uniform distributions, assumed known. From these distribution, we draw $10^5$ binaries and inject the corresponding signals in simulated noise representative of O3 sensitivity.\footnote{The noise curves used are publicly available here: \url{https://dcc.ligo.org/LIGO-T2000012-v1/public}} For each signal, we computed the network signal-to-noise ratio, $\rho$, randomly selected 59 events with $\rho > 10,$ and estimated their parameters using \textsc{Bilby} \citep{ashton:2019} to produce a mock catalogue with properties similar to the BBHs observed during O3. The selection effects can therefore accounted for using the search sensitivity estimates for O3 \citep[][v2]{sensitivityestimate:2023} released along with the third Gravitational Wave Transient Catalog (GWTC-3).

\begin{figure*}
    \centering
    \subfigure[Posterior distribution]{
    \includegraphics[width=0.93\columnwidth]{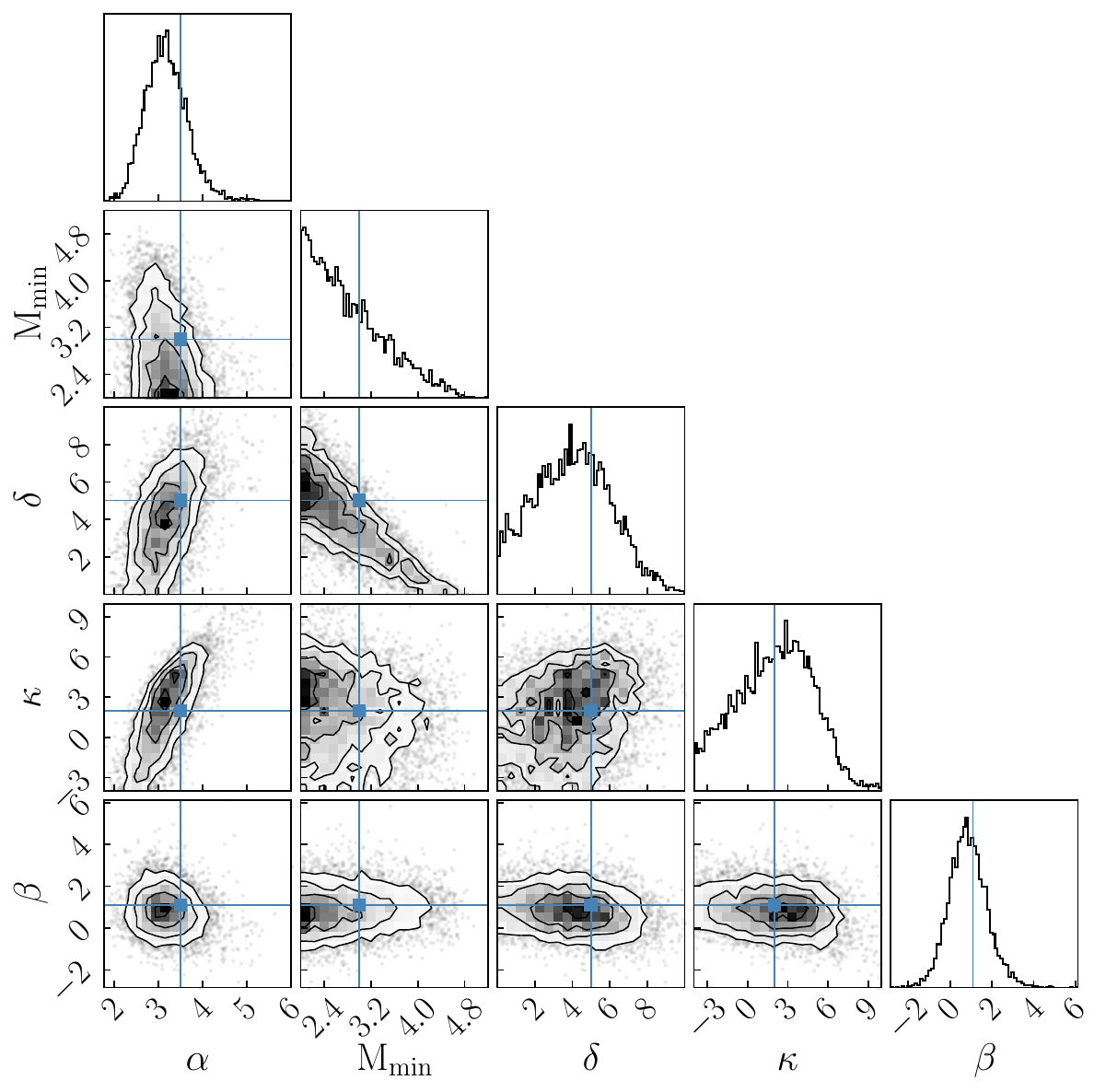}\label{fig:3d_corner}
    }
    \subfigure[Non-parametric reconstruction]{
    \includegraphics[width=0.93\columnwidth]{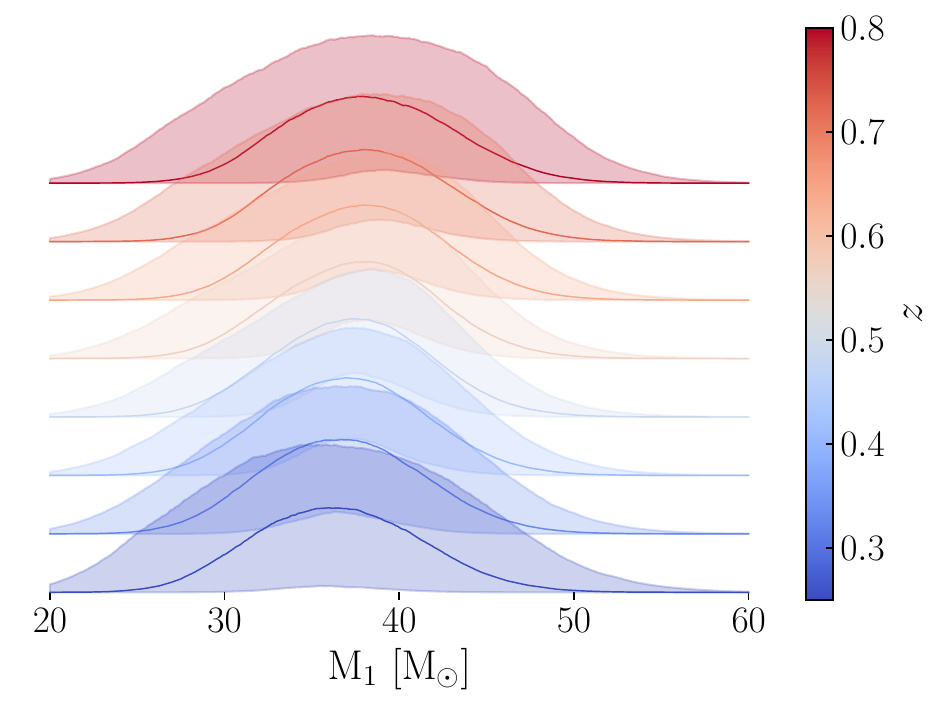}\label{fig:3d_joyplot}
    }
    \caption{Inferred distributions for three-dimensional mock catalogue presented in Section~\ref{sec:3dsim}. Left: Posterior distribution for $\Lambda_\mathrm{TPL}$. The blue cross-hairs mark the true values. Right: Non-parametric marginal $\mathrm{M}_1-z$ distribution. For each redshift value, the mass distribution has been normalised, and the shaded areas mark the $68\%$ credible regions.}
\end{figure*}

We analysed this mock catalogue using the tapered power law (TPL) of the \plpeak~model augmented with (H)DPGMM. The posterior distribution for the parameters $\Lambda_\mathrm{TPL} = \{\alpha,\delta,\mathrm{M}_\mathrm{min},\kappa,\beta\}$ is reported in Figure~\ref{fig:3d_corner}, whereas Figure~\ref{fig:3d_joyplot} shows the non-parametric reconstruction of the marginal $\mathrm{M}_1-z$ distribution. The inferred distributions are consistent with the expectations, both being in agreement with the simulated values and highlighting the presence of the Gaussian feature at around $35\ \msun$. 

\begin{figure}
        \centering
        \includegraphics[width=\columnwidth]{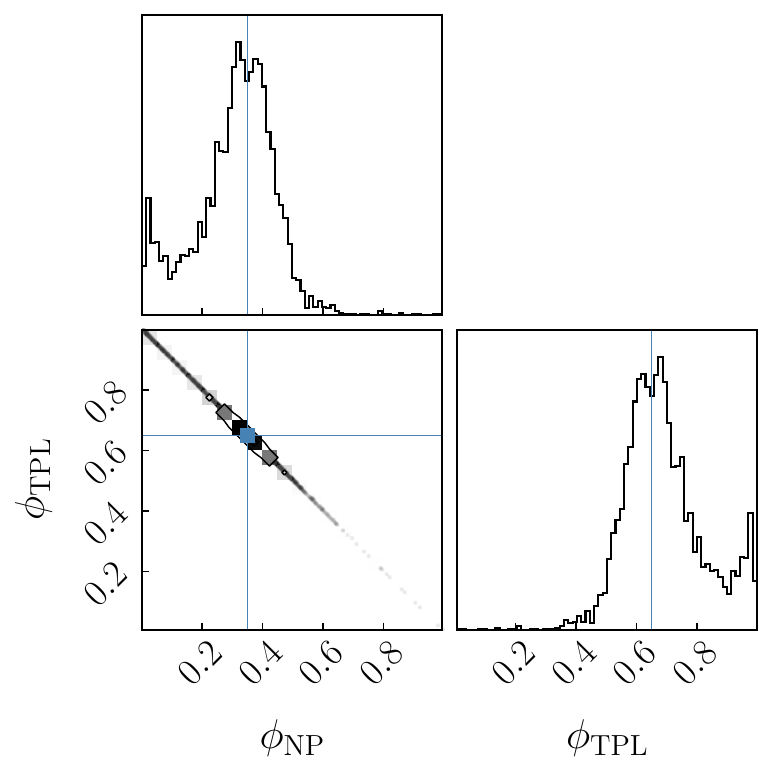}
        \caption{Posterior distribution for $\boldsymbol{\phi}$ using three-dimensional mock catalogue of Section~\ref{sec:3dsim}. The blue cross-hairs mark the true values.}\label{fig:3d_corner_obs_weights}
\end{figure}

The interpretation of the relative weights $(w_\mathrm{NP}, w_\mathrm{TPL})$ requires some care, however, due to the presence of a censored area in the binary parameter space -- namely the low-mass, high-redshift region. Since the GW detectors are not able to detect gravitational signals with such properties, the inferred population distribution will have to rely on extrapolation to cover that specific region of the binary parameter space. The non-parametric methods, however, are by construction informed only by the available data, and they are thus unable to extrapolate beyond the point where data are observable: the detector horizon. Reconstructing the intrinsic distribution in censored areas would yield diverging uncertainties (see, e.g. the yellow shaded area in Figure 2 of \citet{toubiana:2025}, where this point is further discussed).\footnote{A simple way of picturing this is saying that the non-parametric method being `unable to decide' whether the lack of observations in an empty area of the binary parameter space is either due to the intrinsic distribution not having support in such an area or the selection function completely depleting it.} To prevent this issue, \textsc{figaro} is coded such that the inferred distribution vanishes in the censored areas, effectively preventing extrapolation towards high redshifts. In practice this means that, out of all the possible Gaussian components represented by combinations of $\mu_k$ and $\sigma_k$, the ones for which the individual detectability fraction defined in Eq.~\eqref{eq:detectability_def} is smaller than a threshold value, $\xi(\mu_k,\sigma_k) < \xi_\mathrm{th}$, are not considered for inclusion in the mixture. For the examples presented in this paper, we set $\xi_\mathrm{th} = 10^{-3}$. This is by all means an effective artificial cut-off of the inferred distribution in the parameter space and must be kept in mind while interpreting the outcome of the inference. This model-specific prescription, limiting the number of high-redshift binaries as a result of the exclusion of the corresponding mixture components, affects the calculation of the overall detectability fraction, $\xi(\Theta)$, skewing the inference of $w_\mathrm{NP}$ towards smaller values. In such cases, it can become difficult to distinguish whether the presence of the non-parametric channel is required by the data or not. The posterior distribution for the observed mixture fraction $\boldsymbol{\phi}$, reported in Figure~\ref{fig:3d_corner_obs_weights}, solves the issue showing that the non-parametric channel is actually required to account for $\sim 35\%$ of the observed GW events.

The intrinsic mixture fractions, $\mathbf{w,}$ are inherently affected by the specific choices made in the non-parametric model to cure the divergences due to the presence of censored regions in the parameter space. Therefore, while $\mathbf{w}$ is useful in the study of mixtures of parametric models only, in the specific context of AMMs we believe that the main quantities to consider to assess the presence (or absence) of some features in the data are the observed mixture fractions, $\boldsymbol{\phi}$. These quantities are exclusively driven by the number of observed events associated with each mixture component, as shown in Eq.~\eqref{eq:phi_dirichletdist}, and as such the most robust against the effects of selection-bias deconvolution.

\section{BBHs from GWTC-3}\label{sec:o3data}
Having demonstrated the capability of our approach in simulated scenarios, we now analyse the primary mass, mass ratio, and redshift distribution of BBHs using the publicly available GW events detected by the LVK collaboration. In this section, we describe how we made use of the 69 GW events released in GWTC-3 \citep{gwtc3:2023} with false-alarm rates of < 1 yr$^{-1}$. The selection effects were accounted for using the sensitivity estimates for the first three observing runs \citep[O1+O2+O3 --][v2]{sensitivityestimate_full:2023} released along with GWTC-3. This choice was driven by the fact that the focus of this work is on the new methodology that we introduce rather than on the astrophysical interpretation of the findings. An application of this method to the newly released GWTC-4.0 and a discussion of the results yielded will be the subject of a future paper.

\subsection{Parametrised model}\label{sec:o3parametric}
In this first analysis, we applied the same AMM used in Section~\ref{sec:3dsim}, TPL+NP, with $\Lambda_\mathrm{TPL} = \{\alpha,\delta,\mathrm{M}_\mathrm{min},\kappa,\beta\}$. $\mathrm{M}_\mathrm{max}$ is fixed at 100 $\msun$ to ensure that the parametric distribution has non-zero support on the whole mass range. This analysis has a similar concept to the one presented in \citet{rinaldi:2025:features}, where we analysed the same dataset using a weighted superposition of a TPL and a Gaussian peak with independent redshift evolutions; following this paper, we included only the primary mass, mass ratio, and redshift in the analysis, and we assumed a fixed isotropic spin distribution (see Section~3 of \citealp{rinaldi:2025:features}). Moreover, we compare our findings with the ones presented in the aforementioned paper.

\begin{figure}
        \centering
        \includegraphics[width=\columnwidth]{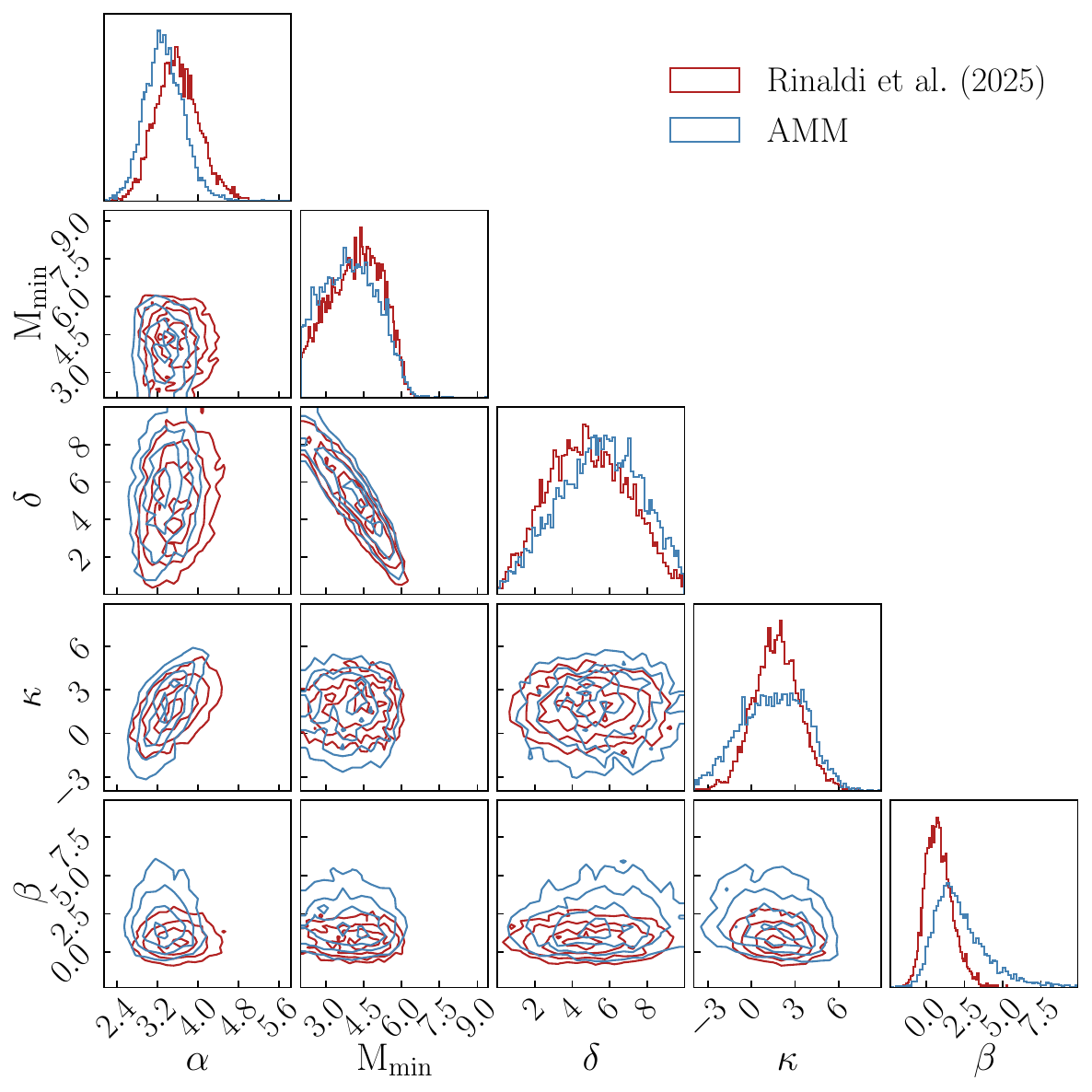}
        \caption{Posterior distribution for $\Lambda_\mathrm{TPL}$ using GWTC-3.}\label{fig:gwtc3_post_lambda}
\end{figure}

\begin{figure*}
    \centering
    \subfigure[Non-parametric reconstruction]{
    \includegraphics[width=0.93\columnwidth]{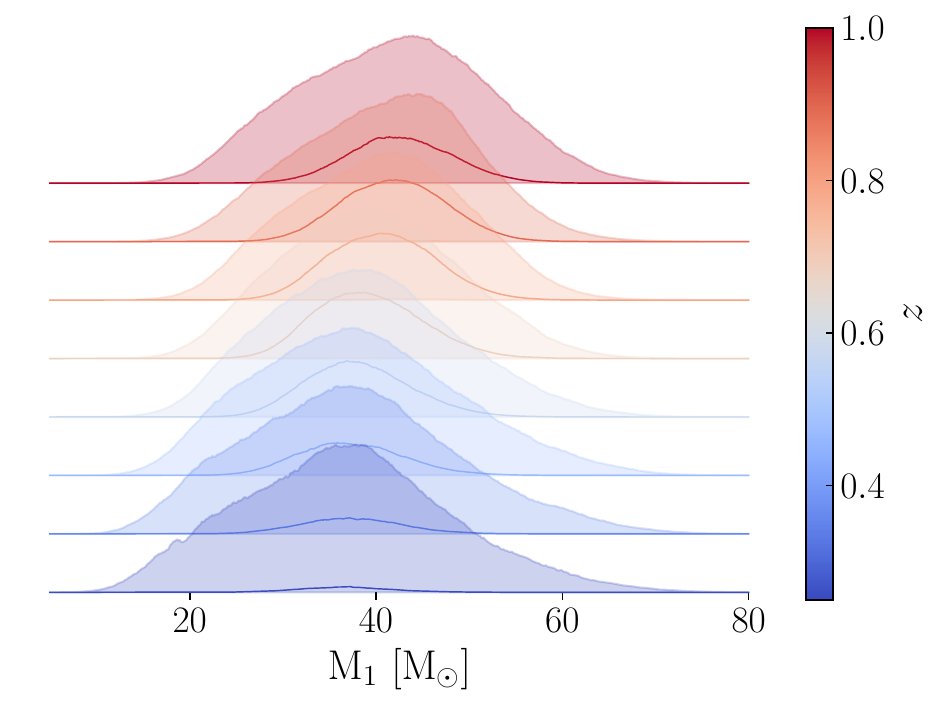}\label{fig:gwtc3_nonpar}
    }
    \subfigure[Remapped posterior]{\includegraphics[width=0.93\columnwidth]{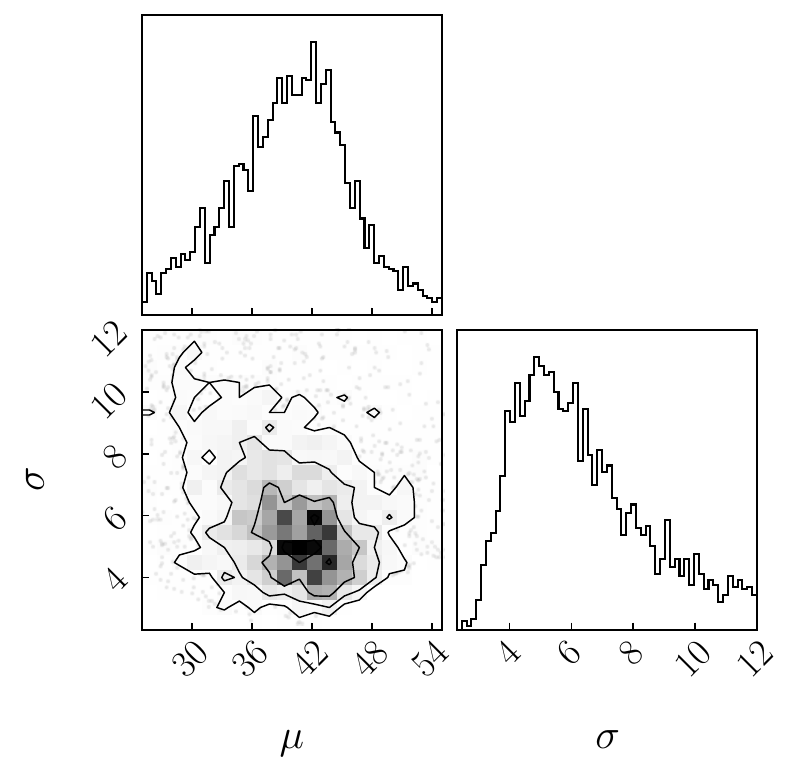}\label{fig:gwtc3_remapped}
    }
    \caption{Left: $\mathrm{M}_1-z$ non-parametric reconstruction using GWTC-3 events. For each redshift value, the mass distribution has been normalised, and the shaded areas mark the $68\%$ credible regions. Right: Posterior distribution for $\mu$ and $\sigma$ of a Gaussian distribution obtained from the non-parametric reconstruction using the remapping procedure presented in \citet{rinaldi:2025:np2p}.}
\end{figure*}

The posterior distribution for $\Lambda_\mathrm{TPL}$ is reported in Figure~\ref{fig:gwtc3_post_lambda}, which is found to be in good agreement with \citet{rinaldi:2025:features}. The differences between the two distributions are to be ascribed to the fact that the NP channel used in this analysis has more flexibility than the Gaussian peak, thus affecting the inference of the TPL parameters differently. The reconstructed marginal non-parametric distribution for $\mathrm{M}_1$ and $z$ is reported in Figure~\ref{fig:gwtc3_nonpar}, highlighting the presence of the $\sim 35\ \msun$ pileup. This non-parametric reconstruction can be converted into a posterior on the parameters of a Gaussian distribution using the remapping procedure presented in \citet{rinaldi:2025:np2p}: the result of this remapping is shown in Figure~\ref{fig:gwtc3_remapped}. We find $\mu = 40^{+7}_{-8}\ \msun$ and $\sigma = 6^{+4}_{-2}\ \msun$, which is in agreement with the expectation of $\mu\sim 35\ \msun$ and $\sigma \sim 4\ \msun$ \citep{astrodistGWTC3:2023}. The inferred relative weight of the NP channel is $w_\mathrm{NP} = 0.003^{+0.2}_{-0.003}$, corresponding to an observed relative weight $\phi_\mathrm{NP} = 0.20^{+0.08}_{-0.11}$ -- 13 GW events associated with the non-parametric channel.

\subsection{Population synthesis}\label{sec:o3popsynth}
In this last example, we will show that astrophysical models can also be augmented with non-parametric methods. We consider an isolated evolution channel using the rapid binary population synthesis code \textsc{sevn}\footnote{Publicly available at \url{https://gitlab.com/sevncodes/sevn}} \citep{spera:2019,mapelli:2020,iorio:2023}, and in particular the publicly available catalogues\footnote{\url{https://zenodo.org/records/7794546}} \citep[][v2]{sevncat:2023} released alongside \citet{iorio:2023}. For simplicity, we made use of the Fiducial model assuming only one metallicity ($Z = 10^{-3}$) and one value for the common envelope efficiency ($\alpha_\mathrm{CE} = 3$). This model is described in detail in Section~3.2 of \citet{iorio:2023}. To obtain a probability density function, we fitted a GMM approximant to the the $\mathrm{M}_1$ and $q$ SEVN samples.
Other choices for the approximant are also possible, such as the normalising flow emulator presented in \citet{colloms:2025}, which allows for the exploration of different values for $\alpha_\mathrm{CE}$ and $Z$. The redshift model is the one given in Eq.~\eqref{eq:redshiftdist}; therefore, in this case, $\Lambda_\mathrm{SEVN} = \{\kappa\}$.

\begin{figure*}
    \centering
    \subfigure[Non-parametric reconstructions]{\includegraphics[width=0.93\columnwidth]{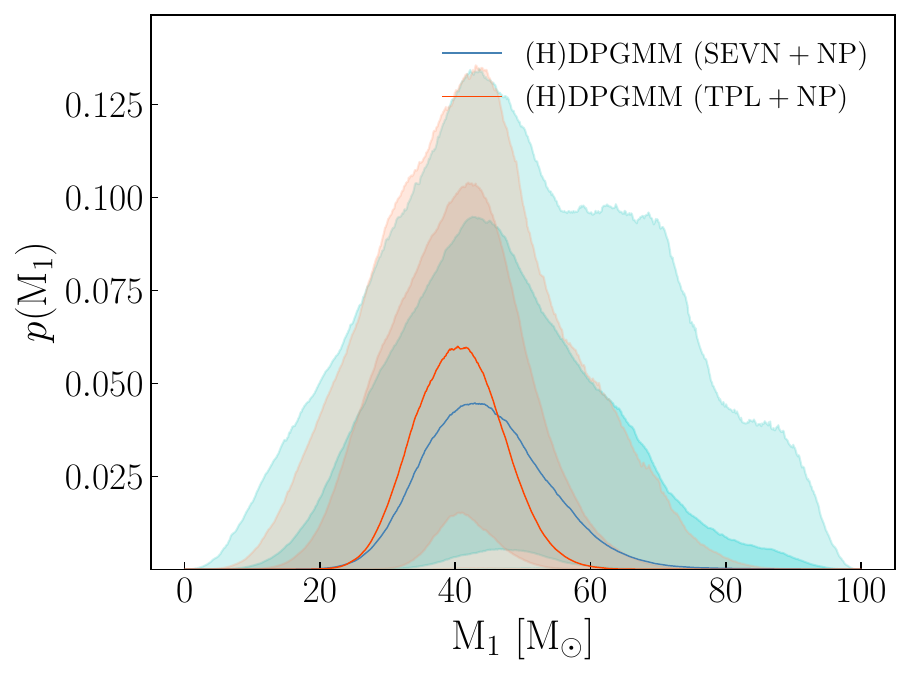}\label{fig:sevn_nonpar}
    }
    \subfigure[Astrophysical/parametric models]{\includegraphics[width=0.93\columnwidth]{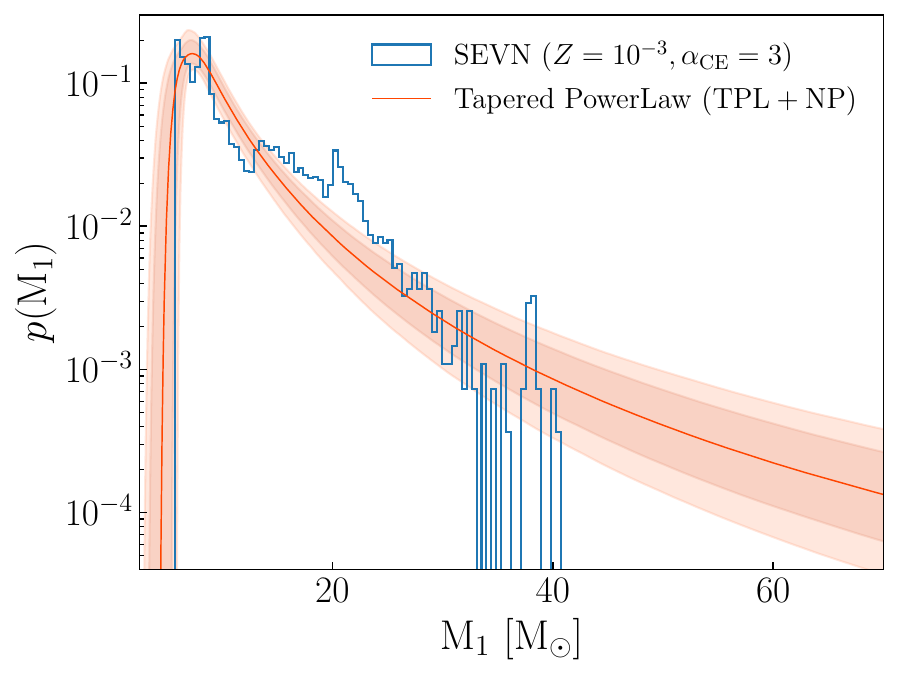}\label{fig:sevn_informed}}
    \caption{Left: Marginal non-parametric reconstruction for $\mathrm{M}_1$ in the SEVN+NP case (blue) and TPL+NP case (red). Right: Comparison between SEVN catalogue used in Section~\ref{sec:o3popsynth} (blue histogram) and the TPL inferred in Section~\ref{sec:o3parametric} (solid red). In both panels, the shaded areas represent the $68\%$ and $90\%$ credible regions.}
\end{figure*}

We find that, for the SEVN+NP model, $\kappa = 4.2^{+2.5}_{-3.9}$. Concerning the relative importance of the two channels, we obtain $w_\mathrm{NP} = 0.01^{+0.1}_{-0.01}$, corresponding to an observed relative weight of $\phi_\mathrm{NP} = 0.35\pm 0.14$. This suggests that around 44 out of 69 GW events can be explained using the isolated evolution model, whereas the remaining ones have to be accounted for using the non-parametric channel.

When comparing the primary-mass non-parametric reconstructions of the SEVN+NP and the TPL+NP models, reported in Figure~\ref{fig:sevn_nonpar}, we see that the former has more support towards more massive BHs. This is a direct consequence of the difference in support between the PL model -- which is deliberately extended all the way to $\mathrm{M}_\mathrm{max} = 100\ \msun$ -- and the SEVN catalogue, which is cut at around $40\ \msun$ due to the pair-instability model used (both reported in Figure~\ref{fig:sevn_informed}). This highlights the fact that not only is the isolated evolution model used in this section incapable of describing the $\sim 35\ \msun$ feature, some additional channels (i.e. dynamical models) are needed to explain the high-mass end of the spectrum. We remind the reader, however, that the example presented here is purely for demonstration purposes: an in-depth study investigating the features that population synthesis codes are able or unable to predict is beyond the scope of this work and will be the subject of a future study.

\section{Summary}\label{sec:conclusions}
In this paper, we introduce the concept of the AMM, a model designed to allow parametric and astrophysical models to accommodate unforeseen features in the analysed data. We achieved this by building a weighted superposition of parametric models and an additional non-parametric model. We validated our formalism applying it to the reconstruction of simulated datasets mimicking the currently available GW observations. We also analysed the real GW events from GWTC-3, finding good agreement with the available literature about the most pronounced features of the primary mass distribution.

The possibility of equipping astrophysical models with non-parametric methods will be extremely valuable to both data analysts and theoreticians. Having a channel able to collect events that are unlikely to be explained by the physically informed models included in the analysis will prevent biases due to mismodelling, as well as highlighting the areas in which astrophysical models need to be further developed. With the grand total of observed GWs after the first third of O4, just above the 150 detection, the newly released GWTC-4.0 will be a perfect playground for AMMs to possibly reveal new features in the astrophysical distribution of BBHs while strengthening our understanding of the formation paths of the compact objects that populate our Universe. 

\begin{acknowledgements}
The author is grateful to Walter~Del~Pozzo, Gabriele~Demasi, and Michela~Mapelli for useful discussion, to Vasco~Gennari, Giuliano~Iorio, and Amedeo~Romagnolo for comments on the initial draft of this paper, and to Gabriele also for providing the script used to generate the mock data used in Section~\ref{sec:3dsim}.
This work was funded by the Deut\-sche For\-schungs\-ge\-mein\-schaft (DFG, German Research Foundation) – project number 546677095. The author acknowledges financial support from the German Excellence Strategy via the Heidelberg Cluster of Excellence (EXC 2181 - 390900948) STRUCTURES, from the state of Baden-W\"urttemberg through bwHPC, from the German Research Foundation (DFG) through grants INST 35/1597-1 FUGG and INST 35/1503-1 FUGG and from the European Research Council for the ERC Consolidator grant DEMOBLACK, under contract no. 770017.

This research has made use of data or software obtained from the Gravitational Wave Open Science Center (gwosc.org), a service of the LIGO Scientific Collaboration, the Virgo Collaboration, and KAGRA. This material is based upon work supported by NSF's LIGO Laboratory which is a major facility fully funded by the National Science Foundation, as well as the Science and Technology Facilities Council (STFC) of the United Kingdom, the Max-Planck-Society (MPS), and the State of Niedersachsen/Germany for support of the construction of Advanced LIGO and construction and operation of the GEO600 detector. Additional support for Advanced LIGO was provided by the Australian Research Council. Virgo is funded, through the European Gravitational Observatory (EGO), by the French Centre National de Recherche Scientifique (CNRS), the Italian Istituto Nazionale di Fisica Nucleare (INFN) and the Dutch Nikhef, with contributions by institutions from Belgium, Germany, Greece, Hungary, Ireland, Japan, Monaco, Poland, Portugal, Spain. KAGRA is supported by Ministry of Education, Culture, Sports, Science and Technology (MEXT), Japan Society for the Promotion of Science (JSPS) in Japan; National Research Foundation (NRF) and Ministry of Science and ICT (MSIT) in Korea; Academia Sinica (AS) and National Science and Technology Council (NSTC) in Taiwan.
\end{acknowledgements}
\bibliographystyle{aa}
\bibliography{bibliography.bib}

@ARTICLE{mandel:2018,
       author = {{Mandel}, Ilya and {Farr}, Will M. and {Gair}, Jonathan R.},
        title = "{Extracting distribution parameters from multiple uncertain observations with selection biases}",
      journal = {\mnras},
         year = 2019,
        month = jun,
       volume = {486},
       number = {1},
        pages = {1086-1093},
          doi = {10.1093/mnras/stz896},
archivePrefix = {arXiv},
       eprint = {1809.02063},
 primaryClass = {physics.data-an},
       adsurl = {https://ui.adsabs.harvard.edu/abs/2019MNRAS.486.1086M}
}

@INCOLLECTION{vitale:2022,
       author = {{Vitale}, Salvatore and {Gerosa}, Davide and {Farr}, Will M. and {Taylor}, Stephen R.},
        title = "{Inferring the Properties of a Population of Compact Binaries in Presence of Selection Effects}",
    booktitle = {Handbook of Gravitational Wave Astronomy},
         year = 2022,
       editor = {{Bambi}, Cosimo and {Katsanevas}, Stavros and {Kokkotas}, Konstantinos D.},
          eid = {45},
        pages = {45},
          doi = {10.1007/978-981-15-4702-7_45-1},
       adsurl = {https://ui.adsabs.harvard.edu/abs/2022hgwa.bookE..45V}
}

@ARTICLE{gwtc3:2023,
       author = {{Abbott}, R. and {Abbott}, T.~D. and {Acernese}, F. and {Ackley}, K. and {Adams}, C. and {Adhikari}, N. and {Adhikari}, R.~X. and {Adya}, V.~B. and {Affeldt}, C. and {Agarwal}, D. and {Agathos}, M. and {Agatsuma}, K. and {Aggarwal}, N. and {Aguiar}, O.~D. and others},
        title = "{GWTC-3: Compact Binary Coalescences Observed by LIGO and Virgo during the Second Part of the Third Observing Run}",
      journal = {PRX},
         year = 2023,
        month = oct,
       volume = {13},
       number = {4},
          eid = {041039},
        pages = {041039},
          doi = {10.1103/PhysRevX.13.041039},
archivePrefix = {arXiv},
       eprint = {2111.03606},
 primaryClass = {gr-qc},
       adsurl = {https://ui.adsabs.harvard.edu/abs/2023PhRvX..13d1039A}
}

@article{ligodetector:2015,
    author = {{Aasi}, J. and {Abbott}, B. P. and {Abbott}, R. and others},
    collaboration = "LIGO Scientific",
    title = "{Advanced LIGO}",
    eprint = "1411.4547",
    archivePrefix = "arXiv",
    primaryClass = "gr-qc",
    doi = "10.1088/0264-9381/32/7/074001",
    journal = "CQG",
    volume = "32",
    pages = "074001",
    year = "2015"
}

@article{virgodetector:2015,
    author = {{Acernese}, F. and {Agathos}, M. and {Agatsuma}, K. and others},
    collaboration = "VIRGO",
    title = "{Advanced Virgo: a second-generation interferometric gravitational wave detector}",
    eprint = "1408.3978",
    archivePrefix = "arXiv",
    primaryClass = "gr-qc",
    doi = "10.1088/0264-9381/32/2/024001",
    journal = "CQG",
    volume = "32",
    number = "2",
    pages = "024001",
    year = "2015"
}

@article{kagradetector:2021,
    author = {{Akutsu}, T. and others},
    title = "{Overview of KAGRA: Detector design and construction history}",
    journal = {PTEP},
    volume = {2021},
    number = {5},
    pages = {05A101},
    year = {2020},
    month = {08},
    issn = {2050-3911},
    doi = {10.1093/ptep/ptaa125},
    url = {https://doi.org/10.1093/ptep/ptaa125},
    eprint = {https://academic.oup.com/ptep/article-pdf/2021/5/05A101/37974994/ptaa125.pdf},
}

@ARTICLE{belczynski:2016,
       author = {{Belczynski}, K. and others},
        title = "{The effect of pair-instability mass loss on black-hole mergers}",
      journal = {\aap},
         year = 2016,
        month = oct,
       volume = {594},
          eid = {A97},
        pages = {A97},
          doi = {10.1051/0004-6361/201628980},
archivePrefix = {arXiv},
       eprint = {1607.03116},
 primaryClass = {astro-ph.HE},
       adsurl = {https://ui.adsabs.harvard.edu/abs/2016A&A...594A..97B}
}

@ARTICLE{vink:2021,
       author = {{Vink}, Jorick S. and {Higgins}, Erin R. and {Sander}, Andreas A.~C. and {Sabhahit}, Gautham N.},
        title = "{Maximum black hole mass across cosmic time}",
      journal = {\mnras},
         year = 2021,
        month = jun,
       volume = {504},
       number = {1},
        pages = {146-154},
          doi = {10.1093/mnras/stab842},
archivePrefix = {arXiv},
       eprint = {2010.11730},
 primaryClass = {astro-ph.HE},
       adsurl = {https://ui.adsabs.harvard.edu/abs/2021MNRAS.504..146V}
}

@ARTICLE{roepke:2023,
       author = {{R{\"o}pke}, Friedrich K. and {De Marco}, Orsola},
        title = "{Simulations of common-envelope evolution in binary stellar systems: physical models and numerical techniques}",
      journal = {LRCA},
         year = 2023,
        month = dec,
       volume = {9},
       number = {1},
          eid = {2},
        pages = {2},
          doi = {10.1007/s41115-023-00017-x},
archivePrefix = {arXiv},
       eprint = {2212.07308},
 primaryClass = {astro-ph.SR},
       adsurl = {https://ui.adsabs.harvard.edu/abs/2023LRCA....9....2R}
}

@ARTICLE{marchant:2021,
       author = {{Marchant}, Pablo and {Pappas}, Kaliro{\"e} M.~W. and {Gallegos-Garcia}, Monica and {Berry}, Christopher P.~L. and {Taam}, Ronald E. and {Kalogera}, Vicky and {Podsiadlowski}, Philipp},
        title = "{The role of mass transfer and common envelope evolution in the formation of merging binary black holes}",
      journal = {\aap},
         year = 2021,
        month = jun,
       volume = {650},
          eid = {A107},
        pages = {A107},
          doi = {10.1051/0004-6361/202039992},
archivePrefix = {arXiv},
       eprint = {2103.09243},
 primaryClass = {astro-ph.SR},
       adsurl = {https://ui.adsabs.harvard.edu/abs/2021A&A...650A.107M}
}

@ARTICLE{gallegos-garcia:2021,
       author = {{Gallegos-Garcia}, Monica and {Berry}, Christopher P.~L. and {Marchant}, Pablo and {Kalogera}, Vicky},
        title = "{Binary Black Hole Formation with Detailed Modeling: Stable Mass Transfer Leads to Lower Merger Rates}",
      journal = {\apj},
         year = 2021,
        month = dec,
       volume = {922},
       number = {2},
          eid = {110},
        pages = {110},
          doi = {10.3847/1538-4357/ac2610},
archivePrefix = {arXiv},
       eprint = {2107.05702},
 primaryClass = {astro-ph.HE},
       adsurl = {https://ui.adsabs.harvard.edu/abs/2021ApJ...922..110G}
}

@ARTICLE{willcox:2023,
       author = {{Willcox}, Reinhold and {MacLeod}, Morgan and {Mandel}, Ilya and {Hirai}, Ryosuke},
        title = "{The Impact of Angular Momentum Loss on the Outcomes of Binary Mass Transfer}",
      journal = {\apj},
         year = 2023,
        month = dec,
       volume = {958},
       number = {2},
          eid = {138},
        pages = {138},
          doi = {10.3847/1538-4357/acffb1},
archivePrefix = {arXiv},
       eprint = {2308.06666},
 primaryClass = {astro-ph.SR},
       adsurl = {https://ui.adsabs.harvard.edu/abs/2023ApJ...958..138W}
}

@article{rodriguez:2016,
  title = {Binary black hole mergers from globular clusters: Masses, merger rates, and the impact of stellar evolution},
  author = {{Rodriguez}, Carl L. and others},
  journal = {\prd},
  volume = {93},
  issue = {8},
  pages = {084029},
  numpages = {21},
  year = {2016},
  month = {04},
  publisher = {American Physical Society},
  doi = {10.1103/PhysRevD.93.084029},
  url = {https://link.aps.org/doi/10.1103/PhysRevD.93.084029}
}

@ARTICLE{ziosi:2014,
       author = {{Ziosi}, Brunetto Marco and {Mapelli}, Michela and {Branchesi}, Marica and {Tormen}, Giuseppe},
        title = "{Dynamics of stellar black holes in young star clusters with different metallicities - II. Black hole-black hole binaries}",
      journal = {\mnras},
         year = 2014,
        month = jul,
       volume = {441},
       number = {4},
        pages = {3703-3717},
          doi = {10.1093/mnras/stu824},
archivePrefix = {arXiv},
       eprint = {1404.7147},
 primaryClass = {astro-ph.GA},
       adsurl = {https://ui.adsabs.harvard.edu/abs/2014MNRAS.441.3703Z}
}

@ARTICLE{kremer:2020,
       author = {{Kremer}, Kyle and {Spera}, Mario and {Becker}, Devin and {Chatterjee}, Sourav and {Di Carlo}, Ugo N. and {Fragione}, Giacomo and {Rodriguez}, Carl L. and {Ye}, Claire S. and {Rasio}, Frederic A.},
        title = "{Populating the Upper Black Hole Mass Gap through Stellar Collisions in Young Star Clusters}",
      journal = {\apj},
         year = 2020,
        month = nov,
       volume = {903},
       number = {1},
          eid = {45},
        pages = {45},
          doi = {10.3847/1538-4357/abb945},
archivePrefix = {arXiv},
       eprint = {2006.10771},
 primaryClass = {astro-ph.HE},
       adsurl = {https://ui.adsabs.harvard.edu/abs/2020ApJ...903...45K}
}

@ARTICLE{banerjee:2022,
       author = {{Banerjee}, Sambaran},
        title = "{Binary black hole mergers from young massive clusters in the pair-instability supernova mass gap}",
      journal = {\aap},
         year = 2022,
        month = sep,
       volume = {665},
          eid = {A20},
        pages = {A20},
          doi = {10.1051/0004-6361/202142331},
archivePrefix = {arXiv},
       eprint = {2109.14612},
 primaryClass = {astro-ph.HE},
       adsurl = {https://ui.adsabs.harvard.edu/abs/2022A&A...665A..20B}
}

@article{mapelli:2020:review,
author={{Mapelli}, Michela},
title="{Binary Black Hole Mergers: Formation and Populations}",
journal={FrASS},
volume={7},
year={2020},
url={https://www.frontiersin.org/articles/10.3389/fspas.2020.00038},
doi={10.3389/fspas.2020.00038},
issn={2296-987X},   
}

@article{mandel:2022,
title = "{Merging stellar-mass binary black holes}",
journal = {PhR},
volume = {955},
pages = {1-24},
year = {2022},
issn = {0370-1573},
doi = {https://doi.org/10.1016/j.physrep.2022.01.003},
url = {https://www.sciencedirect.com/science/article/pii/S0370157322000175},
author = {Ilya {Mandel} and Alison {Farmer}},
}

@article{astrodistGWTC1:2019,
	doi = {10.3847/2041-8213/ab3800},
	url = {https://doi.org/10.3847/2041-8213/ab3800},
	year = 2019,
	month = {sep},
	publisher = {American Astronomical Society},
	volume = {882},
	number = {2},
	pages = {L24},
    title = "{Binary Black Hole Population Properties Inferred from the First and Second Observing Runs of Advanced {LIGO} and Advanced Virgo}",
	journal = {\apj},
	author = {{Abbott}, B. P. and {Abbott}, R. and {Abbott}, T. D. and others}
}

@article{astrodistGWTC2:2021,
	doi = {10.3847/2041-8213/abe949},
	url = {https://doi.org/10.3847/2041-8213/abe949},
	year = 2021,
	month = {may},
	publisher = {American Astronomical Society},
	volume = {913},
	number = {1},
	pages = {L7},
	author = {{Abbott}, R. and {Abbott}, T. D. and {Acernese}, F. and others},
	title = "{Population Properties of Compact Objects from the Second {LIGO}{\textendash}Virgo Gravitational-Wave Transient Catalog}",
	journal = {\apjl},
}

@ARTICLE{astrodistGWTC3:2023,
       author = {{Abbott}, R. and {Abbott}, T. D. and {Acernese}, F. and others},
        title = "{Population of Merging Compact Binaries Inferred Using Gravitational Waves through GWTC-3}",
      journal = {PRX},
         year = 2023,
        month = jan,
       volume = {13},
       number = {1},
          eid = {011048},
        pages = {011048},
          doi = {10.1103/PhysRevX.13.011048},
archivePrefix = {arXiv},
       eprint = {2111.03634},
 primaryClass = {astro-ph.HE},
       adsurl = {https://ui.adsabs.harvard.edu/abs/2023PhRvX..13a1048A}
}

@ARTICLE{callister:2024:review,
       author = {{Callister}, T.~A.},
        title = "{Observed Gravitational-Wave Populations}",
      journal = {arXiv e-prints},
         year = 2024,
        month = oct,
          eid = {arXiv:2410.19145},
        pages = {arXiv:2410.19145},
          doi = {10.48550/arXiv.2410.19145},
archivePrefix = {arXiv},
       eprint = {2410.19145},
 primaryClass = {astro-ph.HE},
       adsurl = {https://ui.adsabs.harvard.edu/abs/2024arXiv241019145C}
}

@ARTICLE{rinaldi:2025:features,
       author = {{Rinaldi}, Stefano and {Liang}, Yajie and {Demasi}, Gabriele and {Mapelli}, Michela and {Del Pozzo}, Walter},
        title = "{Exploration of features in the black hole mass spectrum inspired by non-parametric analyses of gravitational wave observations}",
      journal = {\aap},
         year = 2025,
        month = oct,
       volume = {702},
          eid = {A52},
        pages = {A52},
          doi = {10.1051/0004-6361/202555870},
archivePrefix = {arXiv},
       eprint = {2506.05929},
 primaryClass = {astro-ph.HE},
       adsurl = {https://ui.adsabs.harvard.edu/abs/2025A&A...702A..52R}
}

@article{fishbach:2017,
doi = {10.3847/2041-8213/aa9bf6},
url = {https://dx.doi.org/10.3847/2041-8213/aa9bf6},
year = {2017},
month = {dec},
publisher = {The American Astronomical Society},
volume = {851},
number = {2},
pages = {L25},
author = {Maya Fishbach and Daniel E. Holz},
title = {Where Are LIGO’s Big Black Holes?},
journal = {ApJL},
}

@article{farah:2023,
doi = {10.3847/1538-4357/aced02},
url = {https://dx.doi.org/10.3847/1538-4357/aced02},
year = {2023},
month = {sep},
publisher = {The American Astronomical Society},
volume = {955},
number = {2},
pages = {107},
author = {Amanda M. Farah and Bruce Edelman and Michael Zevin and Maya Fishbach and Jose María Ezquiaga and Ben Farr and Daniel E. Holz},
title = {Things That Might Go Bump in the Night: Assessing Structure in the Binary Black Hole Mass Spectrum},
journal = {ApJ},
}

@article{talbot:2018,
doi = {10.3847/1538-4357/aab34c},
url = {https://dx.doi.org/10.3847/1538-4357/aab34c},
year = {2018},
month = {apr},
publisher = {The American Astronomical Society},
volume = {856},
number = {2},
pages = {173},
author = {Colm Talbot and Eric Thrane},
title = {Measuring the Binary Black Hole Mass Spectrum with an Astrophysically Motivated Parameterization},
journal = {ApJ},
}

@ARTICLE{gennari:2025,
       author = {{Gennari}, Vasco and {Mastrogiovanni}, Simone and {Tamanini}, Nicola and {Marsat}, Sylvain and {Pierra}, Gregoire},
        title = "{Searching for additional structure and redshift evolution in the observed binary black hole population with a parametric time-dependent mass distribution}",
      journal = {\prd},
         year = 2025,
        month = jun,
       volume = {111},
       number = {12},
          eid = {123046},
        pages = {123046},
          doi = {10.1103/ftw9-7xd5},
archivePrefix = {arXiv},
       eprint = {2502.20445},
 primaryClass = {gr-qc},
       adsurl = {https://ui.adsabs.harvard.edu/abs/2025PhRvD.111l3046G}
}

@article{li:2021,
doi = {10.3847/1538-4357/ac0971},
url = {https://dx.doi.org/10.3847/1538-4357/ac0971},
year = {2021},
month = {aug},
publisher = {The American Astronomical Society},
volume = {917},
number = {1},
pages = {33},
author = {Li, Yin-Jie and Wang, Yuan-Zhu and Han, Ming-Zhe and Tang, Shao-Peng and Yuan, Qiang and Fan, Yi-Zhong and Wei, Da-Ming},
title = {A Flexible Gaussian Process Reconstruction Method and the Mass Function of the Coalescing Binary Black Hole Systems},
journal = {ApJ},
}

@article{rinaldi:2022:hdpgmm,
       author = {{Rinaldi}, Stefano and {Del Pozzo}, Walter},
        title = "{(H)DPGMM: a hierarchy of Dirichlet process Gaussian mixture models for the inference of the black hole mass function}",
      journal = {\mnras},
         year = 2022,
        month = feb,
       volume = {509},
       number = {4},
        pages = {5454-5466},
          doi = {10.1093/mnras/stab3224},
archivePrefix = {arXiv},
       eprint = {2109.05960},
 primaryClass = {astro-ph.IM},
       adsurl = {https://ui.adsabs.harvard.edu/abs/2022MNRAS.509.5454R}
}

@ARTICLE{callister:2024,
       author = {{Callister}, Thomas A. and {Farr}, Will M.},
        title = "{Parameter-Free Tour of the Binary Black Hole Population}",
      journal = {PRX},
         year = 2024,
        month = apr,
       volume = {14},
       number = {2},
          eid = {021005},
        pages = {021005},
          doi = {10.1103/PhysRevX.14.021005},
archivePrefix = {arXiv},
       eprint = {2302.07289},
 primaryClass = {astro-ph.HE},
       adsurl = {https://ui.adsabs.harvard.edu/abs/2024PhRvX..14b1005C}
}

@ARTICLE{toubiana:2023,
       author = {{Toubiana}, A. and {Katz}, Michael L. and {Gair}, Jonathan R.},
        title = "{Is there an excess of black holes around 20 M{\ensuremath{\odot}}? Optimizing the complexity of population models with the use of reversible jump MCMC.}",
      journal = {\mnras},
         year = 2023,
        month = oct,
       volume = {524},
       number = {4},
        pages = {5844-5853},
          doi = {10.1093/mnras/stad2215},
archivePrefix = {arXiv},
       eprint = {2305.08909},
 primaryClass = {gr-qc},
       adsurl = {https://ui.adsabs.harvard.edu/abs/2023MNRAS.524.5844T}
}

@ARTICLE{heinzel:2024,
       author = {{Heinzel}, Jack and {Mould}, Matthew and {{\'A}lvarez-L{\'o}pez}, Sof{\'\i}a and {Vitale}, Salvatore},
        title = "{High resolution nonparametric inference of gravitational-wave populations in multiple dimensions}",
      journal = {\prd},
         year = 2025,
        month = mar,
       volume = {111},
       number = {6},
          eid = {063043},
        pages = {063043},
          doi = {10.1103/PhysRevD.111.063043},
archivePrefix = {arXiv},
       eprint = {2406.16813},
 primaryClass = {astro-ph.HE},
       adsurl = {https://ui.adsabs.harvard.edu/abs/2025PhRvD.111f3043H}
}

@ARTICLE{ray:2023,
       author = {{Ray}, Anarya and {Maga{\~n}a Hernandez}, Ignacio and {Mohite}, Siddharth and {Creighton}, Jolien and {Kapadia}, Shasvath},
        title = "{Nonparametric Inference of the Population of Compact Binaries from Gravitational-wave Observations Using Binned Gaussian Processes}",
      journal = {\apj},
         year = 2023,
        month = nov,
       volume = {957},
       number = {1},
          eid = {37},
        pages = {37},
          doi = {10.3847/1538-4357/acf452},
archivePrefix = {arXiv},
       eprint = {2304.08046},
 primaryClass = {gr-qc},
       adsurl = {https://ui.adsabs.harvard.edu/abs/2023ApJ...957...37R}
}

@ARTICLE{rinaldi:2025:np2p,
       author = {{Rinaldi}, Stefano and {Toubiana}, Alexandre and {Gair}, Jonathan R.},
        title = "{Trust the process: mapping data-driven reconstructions to informed models using stochastic processes}",
      journal = {\jcap},
         year = 2025,
        month = dec,
       volume = {2025},
       number = {12},
          eid = {031},
        pages = {031},
          doi = {10.1088/1475-7516/2025/12/031},
archivePrefix = {arXiv},
       eprint = {2506.05153},
 primaryClass = {gr-qc},
       adsurl = {https://ui.adsabs.harvard.edu/abs/2025JCAP...12..031R}
}

@ARTICLE{colloms:2025,
       author = {{Colloms}, Storm and {Berry}, Christopher P.~L. and {Veitch}, John and {Zevin}, Michael},
        title = "{Exploring the Evolution of Gravitational-wave Emitters with Efficient Emulation: Constraining the Origins of Binary Black Holes Using Normalizing Flows}",
      journal = {\apj},
         year = 2025,
        month = aug,
       volume = {988},
       number = {2},
          eid = {189},
        pages = {189},
          doi = {10.3847/1538-4357/ade546},
archivePrefix = {arXiv},
       eprint = {2503.03819},
 primaryClass = {astro-ph.HE},
       adsurl = {https://ui.adsabs.harvard.edu/abs/2025ApJ...988..189C}
}

@ARTICLE{toubiana:2025,
       author = {{Toubiana}, Alexandre and {Gerosa}, Davide and {Mould}, Matthew and {Rinaldi}, Stefano and {Arca Sedda}, Manuel and {Bruel}, Tristan and {Buscicchio}, Riccardo and {Gair}, Jonathan and {Paiella}, Lavinia and {Santoliquido}, Filippo and {Tenorio}, Rodrigo and {Ugolini}, Cristiano},
        title = "{Comparing astrophysical models to gravitational-wave data in the observable space}",
      journal = {arXiv e-prints},
         year = 2025,
        month = jul,
          eid = {arXiv:2507.13249},
        pages = {arXiv:2507.13249},
          doi = {10.48550/arXiv.2507.13249},
archivePrefix = {arXiv},
       eprint = {2507.13249},
 primaryClass = {gr-qc},
       adsurl = {https://ui.adsabs.harvard.edu/abs/2025arXiv250713249T}
}

@article{nguyen:2020,
    author = {{Nguyen}, T. Tin and {Nguyen}, Hien D. and Faicel {Chamroukhi} and Geoffrey J. {McLachlan}},
    editor = {Lishan Liu},
    title = "{Approximation by finite mixtures of continuous density functions that vanish at infinity}",
    journal = {CMS},
    volume = {7},
    number = {1},
    pages = {1750861},
    year  = {2020},
    publisher = {Cogent OA},
    doi = {10.1080/25742558.2020.1750861},
    URL = {https://doi.org/10.1080/25742558.2020.1750861},
    eprint = {https://doi.org/10.1080/25742558.2020.1750861}
}

@ARTICLE{rinaldi:2024:figaro,
       author = {{Rinaldi}, Stefano and {Del Pozzo}, Walter},
        title = "{FIGARO: hierarchical non-parametric inference for population studies}",
      journal = {JOSS},
    publisher = {The Open Journal},
         year = 2024,
        month = may,
       volume = {9},
       number = {97},
        pages = {6589},
          doi = {10.21105/joss.06589},
          url = {https://doi.org/10.21105/joss.06589}
}

@article{ferguson:1973,
author = {{Ferguson}, Thomas S.},
title = {{A Bayesian Analysis of Some Nonparametric Problems}},
volume = {1},
journal = {AnSta},
number = {2},
publisher = {Institute of Mathematical Statistics},
pages = {209 -- 230},
year = {1973},
doi = {10.1214/aos/1176342360},
URL = {https://doi.org/10.1214/aos/1176342360}
}

@ARTICLE{geman:1984,
  author={Geman, Stuart and Geman, Donald},
  journal={ITPAM}, 
  title={Stochastic Relaxation, Gibbs Distributions, and the Bayesian Restoration of Images}, 
  year={1984},
  volume={PAMI-6},
  number={6},
  pages={721-741},
  doi={10.1109/TPAMI.1984.4767596}
}

@article{gelfand:1990,
 ISSN = {01621459},
 URL = {http://www.jstor.org/stable/2289776},
 author = {Alan E. {Gelfand} and Adrian F. M. {Smith}},
 journal = {JASA},
 number = {410},
 pages = {398--409},
 publisher = {[American Statistical Association, Taylor & Francis, Ltd.]},
 title = "{Sampling-Based Approaches to Calculating Marginal Densities}",
 urldate = {2023-08-14},
 volume = {85},
 year = {1990}
}

@article{smith:1993,
 ISSN = {00359246},
 URL = {http://www.jstor.org/stable/2346063},
 author = {A. F. M. {Smith} and G. O. {Roberts}},
 journal = {JRSS},
 number = {1},
 pages = {3--23},
 publisher = {[Royal Statistical Society, Wiley]},
 title = "{Bayesian Computation Via the Gibbs Sampler and Related Markov Chain Monte Carlo Methods}",
 urldate = {2023-08-14},
 volume = {55},
 year = {1993}
}

@article{liu:1994,
 ISSN = {01621459},
 URL = {http://www.jstor.org/stable/2290921},
 author = {Jun S. {Liu}},
 journal = {JASA},
 number = {427},
 pages = {958--966},
 publisher = {[American Statistical Association, Taylor & Francis, Ltd.]},
 title = {The Collapsed Gibbs Sampler in Bayesian Computations with Applications to a Gene Regulation Problem},
 urldate = {2022-08-16},
 volume = {89},
 year = {1994}
}

@article{karnesis:2023,
       author = {{Karnesis}, Nikolaos and {Katz}, Michael L. and {Korsakova}, Natalia and {Gair}, Jonathan R. and {Stergioulas}, Nikolaos},
        title = "{Eryn: a multipurpose sampler for Bayesian inference}",
      journal = {\mnras},
         year = 2023,
        month = dec,
       volume = {526},
       number = {4},
        pages = {4814-4830},
          doi = {10.1093/mnras/stad2939},
archivePrefix = {arXiv},
       eprint = {2303.02164},
 primaryClass = {astro-ph.IM},
       adsurl = {https://ui.adsabs.harvard.edu/abs/2023MNRAS.526.4814K}
}

@software{katz:2023,
  author       = {Michael {Katz} and
                  Nikolaos {Karnesis} and
                  Natalia {Korsakova}},
  title        = {mikekatz04/Eryn: first full release},
  month        = mar,
  year         = 2023,
  publisher    = {Zenodo},
  version      = {v1.0.0},
  doi          = {10.5281/zenodo.7705496},
  url          = {https://doi.org/10.5281/zenodo.7705496}
}

@ARTICLE{foreman-mackey:2013,
       author = {{Foreman-Mackey}, Daniel and {Hogg}, David W. and {Lang}, Dustin and {Goodman}, Jonathan},
        title = "{emcee: The MCMC Hammer}",
      journal = {PASP},
         year = 2013,
        month = mar,
       volume = {125},
       number = {925},
        pages = {306},
          doi = {10.1086/670067},
archivePrefix = {arXiv},
       eprint = {1202.3665},
 primaryClass = {astro-ph.IM},
       adsurl = {https://ui.adsabs.harvard.edu/abs/2013PASP..125..306F}
}

@ARTICLE{veske:2021,
       author = {{Veske}, Do{\u{g}}a and {Bartos}, Imre and {M{\'a}rka}, Zsuzsa and {M{\'a}rka}, Szabolcs},
        title = "{Characterizing the Observation Bias in Gravitational-wave Detections and Finding Structured Population Properties}",
      journal = {\apj},
         year = 2021,
        month = dec,
       volume = {922},
       number = {2},
          eid = {258},
        pages = {258},
          doi = {10.3847/1538-4357/ac27ac},
archivePrefix = {arXiv},
       eprint = {2105.13983},
 primaryClass = {gr-qc},
       adsurl = {https://ui.adsabs.harvard.edu/abs/2021ApJ...922..258V}
}

@ARTICLE{ashton:2019,
       author = {{Ashton}, Gregory and {H{\"u}bner}, Moritz and {Lasky}, Paul D. and {Talbot}, Colm and {Ackley}, Kendall and {Biscoveanu}, Sylvia and {Chu}, Qi and {Divakarla}, Atul and {Easter}, Paul J. and {Goncharov}, Boris and {Hernandez Vivanco}, Francisco and {Harms}, Jan and {Lower}, Marcus E. and {Meadors}, Grant D. and {Melchor}, Denyz and {Payne}, Ethan and {Pitkin}, Matthew D. and {Powell}, Jade and {Sarin}, Nikhil and {Smith}, Rory J.~E. and {Thrane}, Eric},
        title = "{BILBY: A User-friendly Bayesian Inference Library for Gravitational-wave Astronomy}",
      journal = {\apjs},
         year = 2019,
        month = apr,
       volume = {241},
       number = {2},
          eid = {27},
        pages = {27},
          doi = {10.3847/1538-4365/ab06fc},
archivePrefix = {arXiv},
       eprint = {1811.02042},
 primaryClass = {astro-ph.IM},
       adsurl = {https://ui.adsabs.harvard.edu/abs/2019ApJS..241...27A}
}

@misc{sensitivityestimate:2023,
author = {{LIGO Scientific}, Collaboration and {Virgo}, Collaboration and {KAGRA}, Collaboration},
title = {GWTC-3: Compact Binary Coalescences Observed by LIGO and Virgo During the Second Part of the Third Observing Run — O3 Search Sensitivity Estimates},
doi = {10.5281/zenodo.7890437},
howpublished= {\url{https://zenodo.org/records/7890437}},
year = 2023,
}

@misc{sensitivityestimate_full:2023,
author = {{LIGO Scientific}, Collaboration and {Virgo}, Collaboration and {KAGRA}, Collaboration},
title = {GWTC-3: Compact Binary Coalescences Observed by LIGO and Virgo During the Second Part of the Third Observing Run — O1+O2+O3 Search Sensitivity Estimates},
doi = {10.5281/zenodo.7890398},
howpublished= {\url{https://zenodo.org/records/7890398}},
year = 2023,
}

@misc{sevncat:2023,
author = {{Iorio}, Giuliano and {Costa}, Guglielmo and {Mapelli}, Michela and {Spera}, Mario and {Trani}, Alessandro Alberto and {Escobar}, Gaston and {Sgalletta}, Cecilia and {Korb}, Erika and {Santoliquido}, Filippo and {Dall'Amico}, Marco and {Gaspari}, Nicola and {Bressan}, Alessandro},
title = {Dataset from the paper ``Compact object mergers: exploring uncertainties from stellar and binary evolution with SEVN''},
doi = {10.5281/zenodo.7794546},
howpublished= {\url{https://zenodo.org/records/7794546}},
year = 2023,
}

@ARTICLE{iorio:2023,
       author = {{Iorio}, Giuliano and {Mapelli}, Michela and {Costa}, Guglielmo and {Spera}, Mario and {Escobar}, Gast{\'o}n J. and {Sgalletta}, Cecilia and {Trani}, Alessandro A. and {Korb}, Erika and {Santoliquido}, Filippo and {Dall'Amico}, Marco and {Gaspari}, Nicola and {Bressan}, Alessandro},
        title = "{Compact object mergers: exploring uncertainties from stellar and binary evolution with SEVN}",
      journal = {\mnras},
         year = 2023,
        month = sep,
       volume = {524},
       number = {1},
        pages = {426-470},
          doi = {10.1093/mnras/stad1630},
archivePrefix = {arXiv},
       eprint = {2211.11774},
 primaryClass = {astro-ph.HE},
       adsurl = {https://ui.adsabs.harvard.edu/abs/2023MNRAS.524..426I}
}

@ARTICLE{gwtc4:2025,
       author = {{The LIGO Scientific Collaboration} and {The Virgo Collaboration} and {the KAGRA Collaboration}},
        title = "{GWTC-4.0: Updating the Gravitational-Wave Transient Catalog with Observations from the First Part of the Fourth LIGO-Virgo-KAGRA Observing Run}",
      journal = {arXiv e-prints},
         year = 2025,
        month = aug,
          eid = {arXiv:2508.18082},
        pages = {arXiv:2508.18082},
          doi = {10.48550/arXiv.2508.18082},
archivePrefix = {arXiv},
       eprint = {2508.18082},
 primaryClass = {gr-qc},
       adsurl = {https://ui.adsabs.harvard.edu/abs/2025arXiv250818082T}
}

@ARTICLE{mapelli:2020,
       author = {{Mapelli}, Michela and {Spera}, Mario and {Montanari}, Enrico and {Limongi}, Marco and {Chieffi}, Alessandro and {Giacobbo}, Nicola and {Bressan}, Alessandro and {Bouffanais}, Yann},
        title = "{Impact of the Rotation and Compactness of Progenitors on the Mass of Black Holes}",
      journal = {\apj},
         year = 2020,
        month = jan,
       volume = {888},
       number = {2},
          eid = {76},
        pages = {76},
          doi = {10.3847/1538-4357/ab584d},
archivePrefix = {arXiv},
       eprint = {1909.01371},
 primaryClass = {astro-ph.HE},
       adsurl = {https://ui.adsabs.harvard.edu/abs/2020ApJ...888...76M}
}

@ARTICLE{spera:2019,
       author = {{Spera}, Mario and {Mapelli}, Michela and {Giacobbo}, Nicola and {Trani}, Alessandro A. and {Bressan}, Alessandro and {Costa}, Guglielmo},
        title = "{Merging black hole binaries with the SEVN code}",
      journal = {\mnras},
         year = 2019,
        month = may,
       volume = {485},
       number = {1},
        pages = {889-907},
          doi = {10.1093/mnras/stz359},
archivePrefix = {arXiv},
       eprint = {1809.04605},
 primaryClass = {astro-ph.HE},
       adsurl = {https://ui.adsabs.harvard.edu/abs/2019MNRAS.485..889S}
}

@ARTICLE{astrodistGWTC4:2025,
       author = {{The LIGO Scientific Collaboration} and {the Virgo Collaboration} and {the KAGRA Collaboration}},
        title = "{GWTC-4.0: Population Properties of Merging Compact Binaries}",
      journal = {arXiv e-prints},
         year = 2025,
        month = aug,
          eid = {arXiv:2508.18083},
        pages = {arXiv:2508.18083},
          doi = {10.48550/arXiv.2508.18083},
archivePrefix = {arXiv},
       eprint = {2508.18083},
 primaryClass = {astro-ph.HE},
       adsurl = {https://ui.adsabs.harvard.edu/abs/2025arXiv250818083T}
}

@ARTICLE{hirschi:2025,
       author = {{Hirschi}, R. and {Goodman}, K. and {Meynet}, G. and {Maeder}, A. and {Ekstr{\"o}m}, S. and {Eggenberger}, P. and {Georgy}, C. and {Sibony}, Y. and {Yusof}, N. and {Martinet}, S. and {Varma}, Vishnu and {Nomoto}, K.},
        title = "{The fate of rotating massive stars across cosmic times}",
      journal = {\mnras},
         year = 2025,
        month = sep,
          doi = {10.1093/mnras/staf1470},
archivePrefix = {arXiv},
       eprint = {2508.21233},
 primaryClass = {astro-ph.SR},
       adsurl = {https://ui.adsabs.harvard.edu/abs/2025MNRAS.tmp.1413H}
}

@ARTICLE{kruckow:2024,
       author = {{Kruckow}, Matthias U. and {Andrews}, Jeff J. and {Fragos}, Tassos and {Holl}, Berry and {Bavera}, Simone S. and {Briel}, Max and {Gossage}, Seth and {Kovlakas}, Konstantinos and {Rocha}, Kyle A. and {Sun}, Meng and {Srivastava}, Philipp M. and {Xing}, Zepei and {Zapartas}, Emmanouil},
        title = "{The formation of black holes in non-interacting isolated binaries: Gaia black holes as calibrators of stellar winds from massive stars}",
      journal = {\aap},
         year = 2024,
        month = dec,
       volume = {692},
          eid = {A141},
        pages = {A141},
          doi = {10.1051/0004-6361/202452356},
archivePrefix = {arXiv},
       eprint = {2410.18501},
 primaryClass = {astro-ph.SR},
       adsurl = {https://ui.adsabs.harvard.edu/abs/2024A&A...692A.141K}
}

@ARTICLE{romagnolo:2024,
       author = {{Romagnolo}, Amedeo and {Gormaz-Matamala}, Alex C. and {Belczynski}, Krzysztof},
        title = "{On the Maximum Black Hole Mass at Solar Metallicity}",
      journal = {\apjl},
         year = 2024,
        month = apr,
       volume = {964},
       number = {2},
          eid = {L23},
        pages = {L23},
          doi = {10.3847/2041-8213/ad2fbe},
archivePrefix = {arXiv},
       eprint = {2311.18841},
 primaryClass = {astro-ph.SR},
       adsurl = {https://ui.adsabs.harvard.edu/abs/2024ApJ...964L..23R}
}

@ARTICLE{vink:2024,
       author = {{Vink}, Jorick S. and {Sabhahit}, Gautham N. and {Higgins}, Erin R.},
        title = "{The maximum black hole mass at solar metallicity}",
      journal = {\aap},
         year = 2024,
        month = aug,
       volume = {688},
          eid = {L10},
        pages = {L10},
          doi = {10.1051/0004-6361/202450655},
archivePrefix = {arXiv},
       eprint = {2407.07204},
 primaryClass = {astro-ph.SR},
       adsurl = {https://ui.adsabs.harvard.edu/abs/2024A&A...688L..10V}
}

@ARTICLE{merritt:2025,
       author = {{Merritt}, JD and {Stevenson}, Simon and {Sander}, Andreas and {Mandel}, Ilya and {Riley}, Jeff and {Farr}, Ben and {van Son}, L.~A.~C. and {Wagg}, Tom and {Vinciguerra}, Serena and {Jose}, Holden},
        title = "{Implications of modern mass-loss rates for massive stars}",
      journal = {arXiv e-prints},
         year = 2025,
        month = jul,
          eid = {arXiv:2507.17052},
        pages = {arXiv:2507.17052},
          doi = {10.48550/arXiv.2507.17052},
archivePrefix = {arXiv},
       eprint = {2507.17052},
 primaryClass = {astro-ph.SR},
       adsurl = {https://ui.adsabs.harvard.edu/abs/2025arXiv250717052M}
}

@ARTICLE{vanson:2025,
       author = {{van Son}, L.~A.~C. and {Roy}, S.~K. and {Mandel}, I. and {Farr}, W.~M. and {Lam}, A. and {Merritt}, J. and {Broekgaarden}, F.~S. and {Sander}, A.~A.~C. and {Andrews}, J.~J.},
        title = "{Not Just Winds: Why Models Find That Binary Black Hole Formation Is Metallicity-dependent, while Binary Neutron Star Formation Is Not}",
      journal = {\apj},
         year = 2025,
        month = feb,
       volume = {979},
       number = {2},
          eid = {209},
        pages = {209},
          doi = {10.3847/1538-4357/ada14a},
archivePrefix = {arXiv},
       eprint = {2411.02484},
 primaryClass = {astro-ph.HE},
       adsurl = {https://ui.adsabs.harvard.edu/abs/2025ApJ...979..209V}
}
\end{document}